\documentclass[floats,floatfix,twocolumn,nosuperscriptaddress,nofootinbib]{revtex4-1}

\pdfoutput=1
\usepackage{amsfonts,amssymb,amsmath}
\usepackage[usenames,dvipsnames]{xcolor}
\usepackage[unicode,colorlinks=true,linkcolor=blue,urlcolor=blue,citecolor=gray,pdfusetitle]{hyperref}
\usepackage[all]{hypcap}
\usepackage{graphicx}
\usepackage{xspace}
\usepackage[normalem]{ulem}

\usepackage[T1]{fontenc}
\usepackage[utf8]{inputenc}

\usepackage{microtype}

\usepackage{youngtab}

\newcommand{\Caltech}{\affiliation{Theoretical Astrophysics,
    Walter Burke Institute for Theoretical Physics,\\
    California Institute of Technology, Pasadena, California 91125, USA}}

\newcommand{\codename}[1]{\textsc{#1}}

\newcommand{\eps}{\ensuremath{\varepsilon}}

\newcommand{\pd}{\partial}
\newcommand{\cd}{\nabla}

\newcommand{\txt}[1]{{\textrm{\tiny{#1}}}}
\newcommand{\mpl}{\ensuremath{m_\txt{pl}}}
\newcommand{\GR}{\txt{GR}}
\newcommand{\INT}{\txt{int}}
\newcommand{\pont}{{}^{*}\!RR}
\newcommand{\Rf}{ {}^{(4)}\!R}
\newcommand{\Rt}{ {}^{(3)}\!R}

\graphicspath{{Figures/}}

\newcommand{\tabRuns}{%
\begin{table}[b]
\begin{ruledtabular}
  \begin{tabular}{ c | c  c  c  c  c  c  c  c}
    Name & $\frac{m_1}{m_2}$ & $\chi_1$ & $\chi_2 $ &  $\Omega_0 (GM)$ &
    $\frac{t_\mathrm{Merger}}{GM}$ & $\frac{t_\mathrm{RD}}{GM}$ &
    $\frac{m_\mathrm{Final}}{M}$ & $\chi_\mathrm{Final}$ \\
    \hline
    Spin 0.3 & 3.0 & 0.30 & 0.30 & 0.0163 & 5841 &  764 & 0.96 & 0.68\\
    Spin 0.1 & 3.0 & 0.10 & 0.10 & 0.0164  & 5452 & 817 & 0.97 & 0.59\\
    Spin 0.0 & 3.0 & 0.00 & 0.00 & 0.0190  & 3457 & 697 & 0.97 & 0.54\\
  \end{tabular}
\end{ruledtabular}
\caption{%
  Parameters of numerical runs. Each run was performed at low,
  medium, and high resolutions.  We give the mass ratio $m_1/m_2$
  where the subscripts label the black holes.  All of the spins are
  aligned in the $z$-direction, so we give the $\hat z$ component of
  the dimensionless spin vector $\vec{\chi}_A$ for
  each black hole.  The initial orbital frequency is $\Omega_0$.
  Initial orbital parameters were chosen so that the
  eccentricity was below $5 \times 10^{-4}$.  The time
  simulated to merger is $t_\mathrm{Merger}$, and the amount of
  ringdown simulated thereafter is $t_\mathrm{RD}$, both in units of
  $GM$.  The final mass of the remnant black hole is
  $m_\mathrm{Final}$, in units of $M$.  The remnant spins are in the
  $z$-direction, and thus we give the $\hat z$ component
  $\chi_\mathrm{Final}$ of the dimensionless spin.
}
\label{tab:runs}
\end{table}
}

\newcommand{\figWaveformThreeThree}{%
\begin{figure}
  \centering
  \includegraphics[width=\columnwidth]{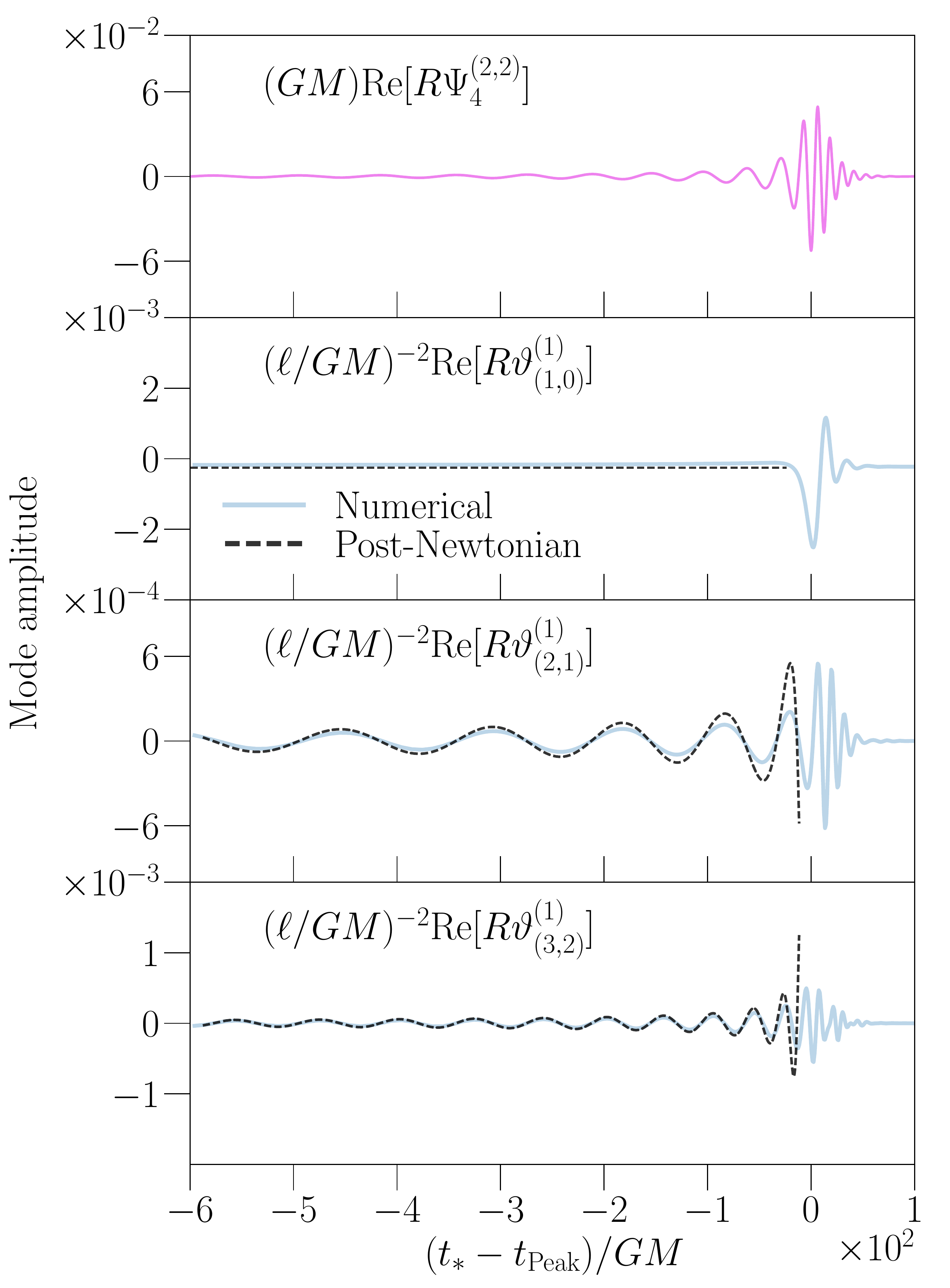}
  \caption{%
    Waveforms for simulation with spin $\chi = 0.3 \hat z$ on each black hole.
    The top panel shows the real part of the $(l=2,m=2)$ mode of the
    spin-weight $-2$ spherical harmonic decomposition of the
    Newman-Penrose scalar $\Psi_4$, extracted at a (large enough) radius of
    $R = 290~GM$. This
    serves as a proxy for the gravitational waveform.  The lower three
    panels show the (1,0), (2,1) and (3,2) scalar spherical harmonic
    modes of the scalar $\vartheta^{(1)}$ at $R = 300~GM$.  The
    numerical values from the simulation are shown by the solid blue
    curves, while the PN calculations are shown by the dashed black
    curves. The time axis corresponds to the approximate retarded time
    (simulation time minus extraction radius) minus the merger time,
    which is computed as the time of peak amplitude of
    $\Psi_4^{(2,2)}$.
  }
  \label{fig:0p3_0p3_Waveform}
\end{figure}
}
\newcommand{\figWaveformOneOne}{%
\begin{figure}
  \centering
  \includegraphics[width=\columnwidth]{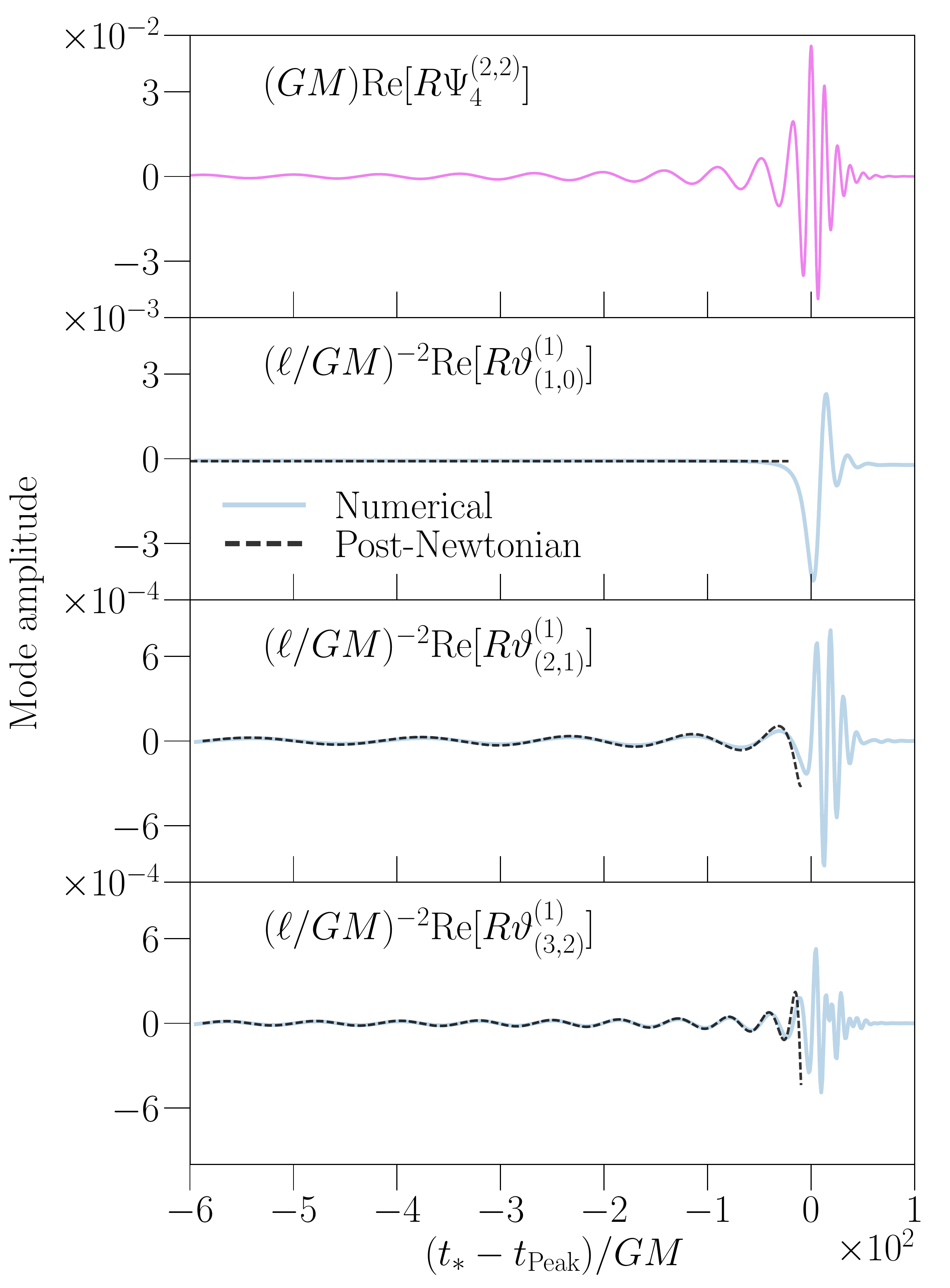}
  \caption{%
    Similar to Fig.~\ref{fig:0p3_0p3_Waveform}, but with
    spin $\chi = 0.1 \hat z$ on each BH.
  }
  \label{fig:0p1_0p1_Waveform}
\end{figure}
}
\newcommand{\figWaveformZeroZero}{%
\begin{figure}
  \centering
  \includegraphics[width=\columnwidth]{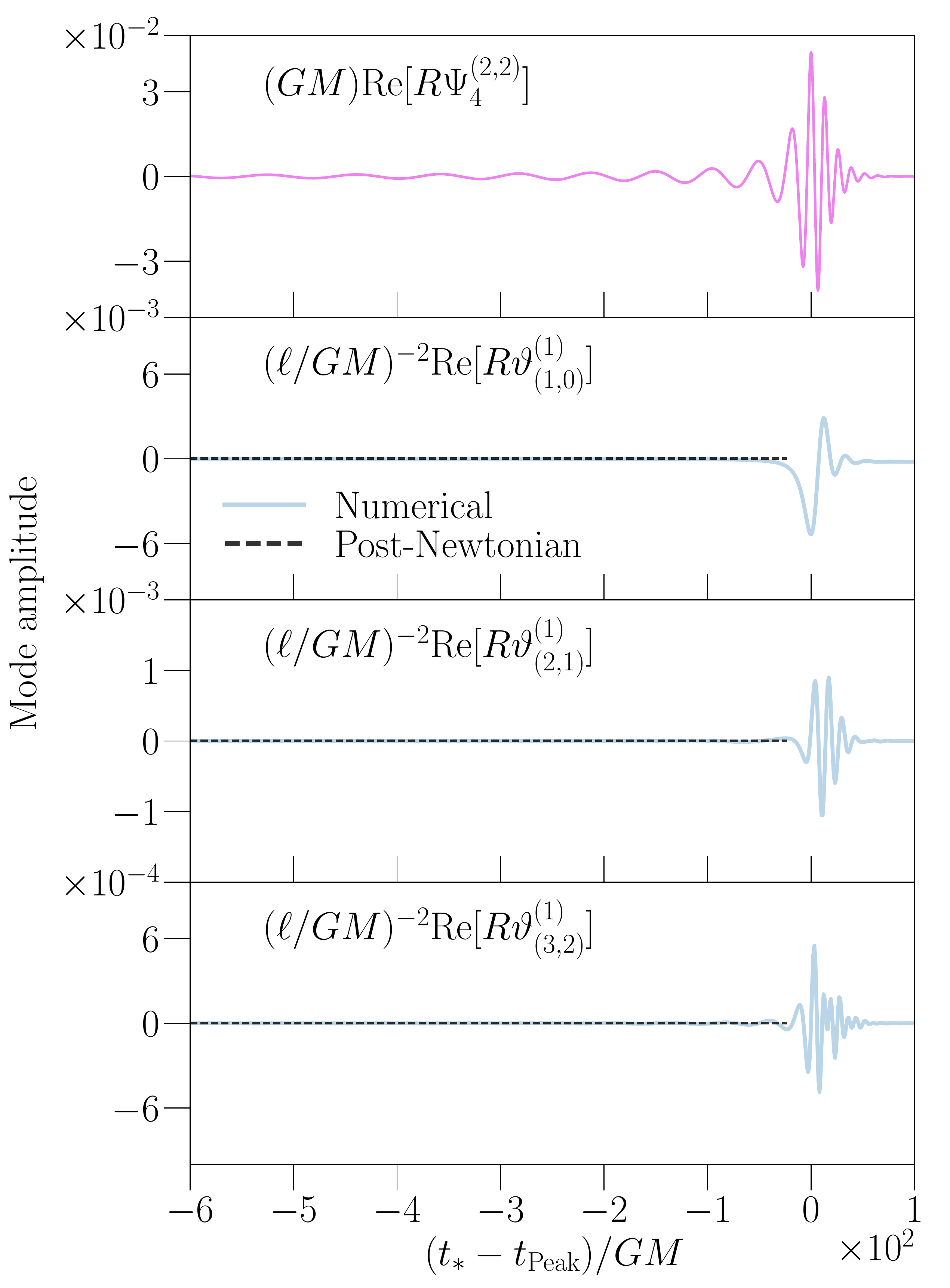}
  \caption{%
    Similar to Fig.~\ref{fig:0p3_0p3_Waveform}, but with
    no spin on either BH.
  }
  \label{fig:0p0_0p0_Waveform}
\end{figure}
}

\newcommand{\figFluxes}{%
\begin{figure}
\centering
\includegraphics[width=\columnwidth]{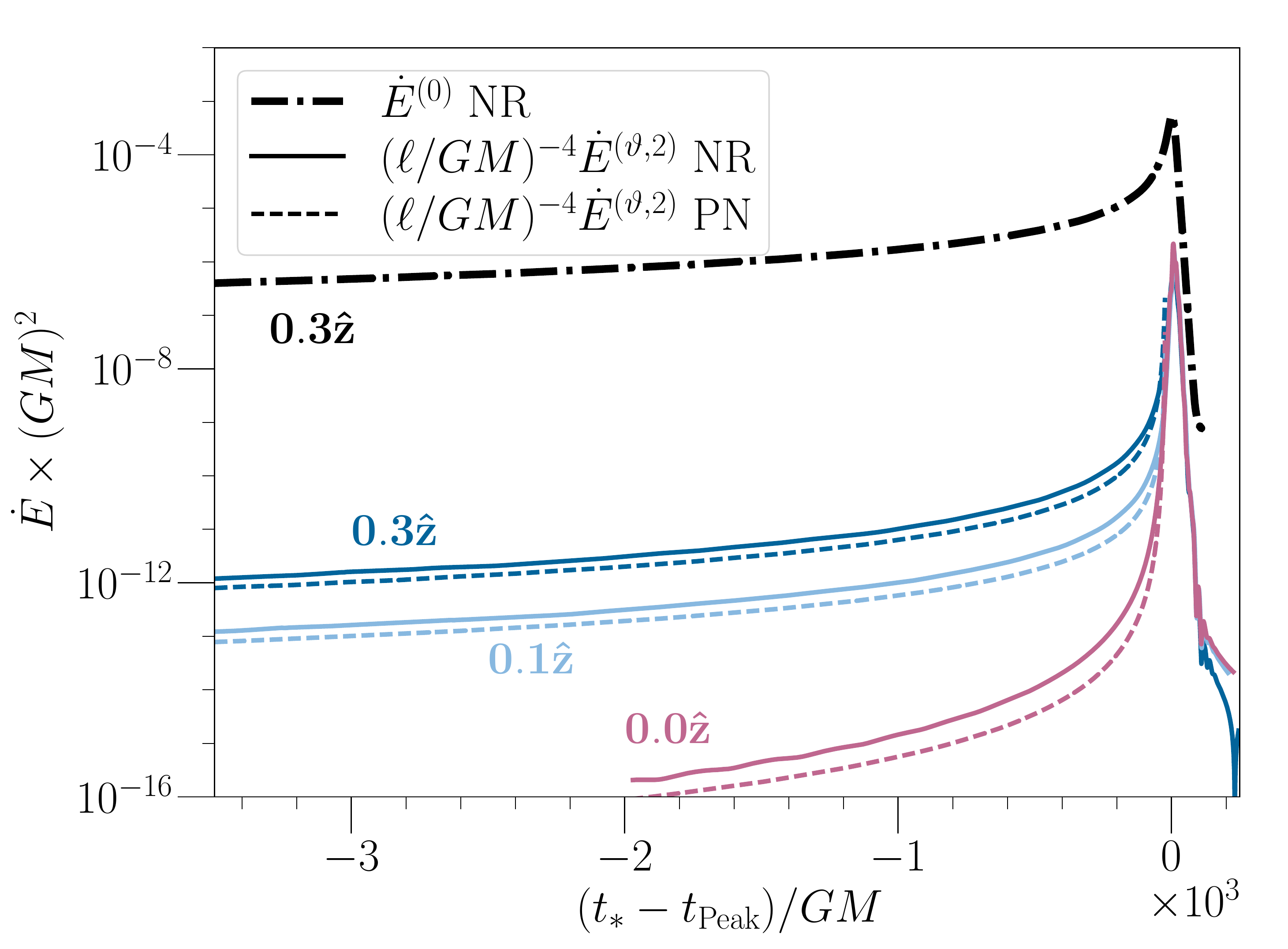}
\caption{%
  Order $(\ell/GM)^0$ and $(\ell/GM)^4$ energy fluxes, as a function
  of time, aligned at the peak of $\Psi_{4}^{(2,2)}$.  We plot the
  order $(\ell/GM)^4$ numerical scalar energy flux extracted at $R =
  300~GM$ [colored solid lines; Eq.~\eqref{eq:NumericalScalarFlux}]
  and the corresponding post-Newtonian approximation [dashed lines,
  Eqs.~\eqref{eq:SpinFluxPN} and \eqref{eq:NoSpinFluxPN}], for the
  highest resolution of each simulation.  We also plot the energy flux at
  order $(\ell/GM)^0$, which consists solely of the background
  gravitational radiation [Eq.~\eqref{eq:GRFlux}], for the spin $0.3$
  simulation (dot-dashed black line); the GW flux is the same order of
  magnitude for all three spin configurations.  The $\mathcal{O}(1)$
  ratio between PN and numerics is likely due to the PN fluxes only
  including $l=2$, whereas numerical quantities are computed with all
  modes up to $l=8$.
}
\label{fig:Fluxes}
\end{figure}
}

\newcommand{\figInstantPert}{%
\begin{figure}
  \centering
  \includegraphics[width=\columnwidth]{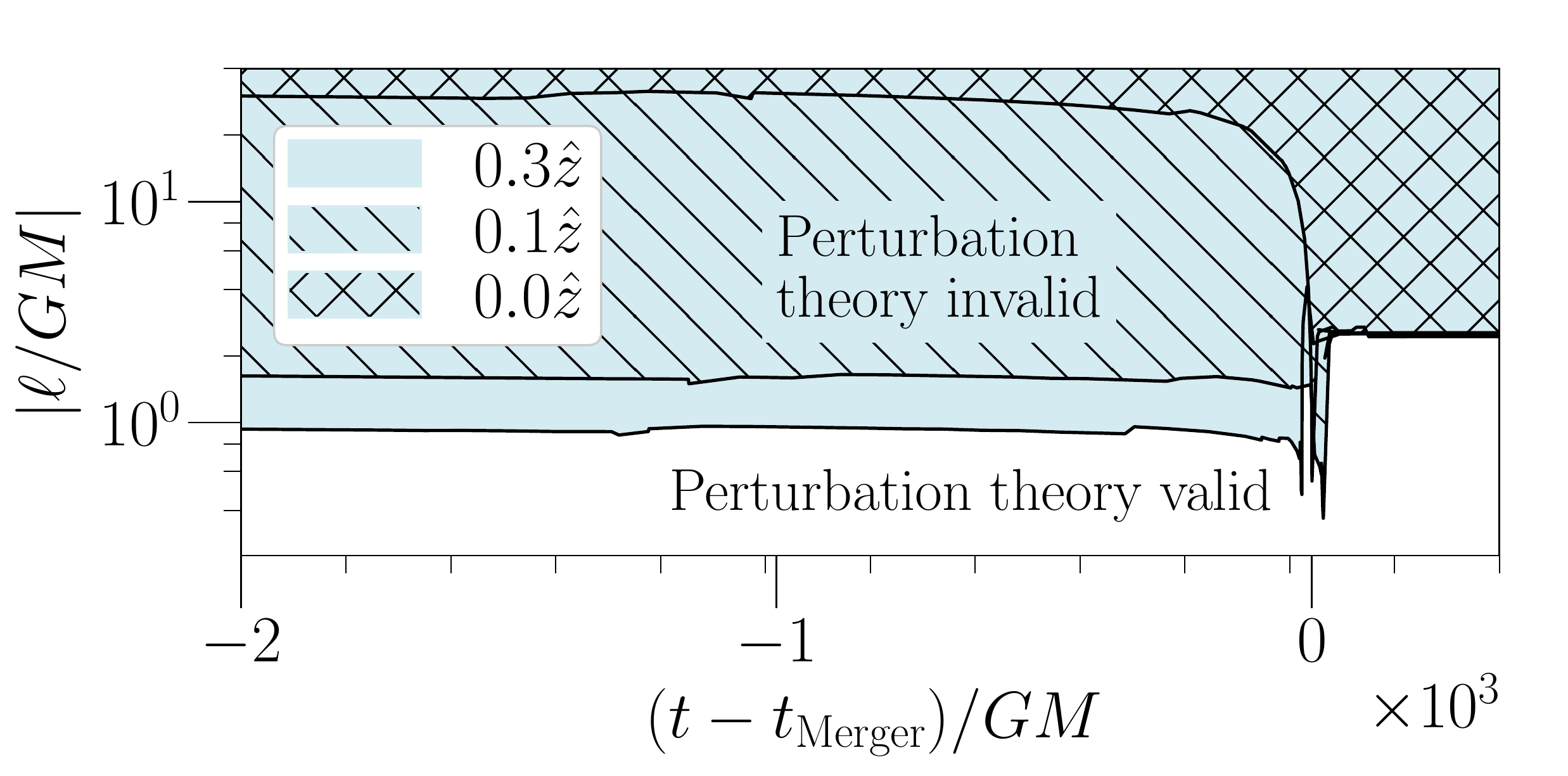}
  \caption{%
    Estimate of instantaneous regime of validity of perturbation
    theory for each of the binary black hole configurations in this
    study, as a function of coordinate time relative to merger.
    Perturbation theory in powers of $|\ell/GM|$ is invalid in the
    shaded region above each curve.  The maximum allowed value of
    $|\ell/GM|$ comes from
    Eq.~\eqref{eq:ell-over-GM-condition-combined}.
    The jaggedness at early times is due to p-refinement of the
    spectral subdomains causing points to cross the mask outside of
    apparent horizons.
    The jump near time of merger is due to formation of the common horizon.
    After merger, the remnant black hole governs
    $|\ell/GM|_{\max}$.  Since all simulations have comparable remnant spins
    (see Table~\ref{tab:runs}), the final values of valid $|\ell/GM|$
    are similar.
  }
  \label{fig:Epsilon}
\end{figure}
}

\newcommand{\figDephasingWindow}{%
\begin{figure}
  \centering
  \includegraphics[width=\columnwidth]{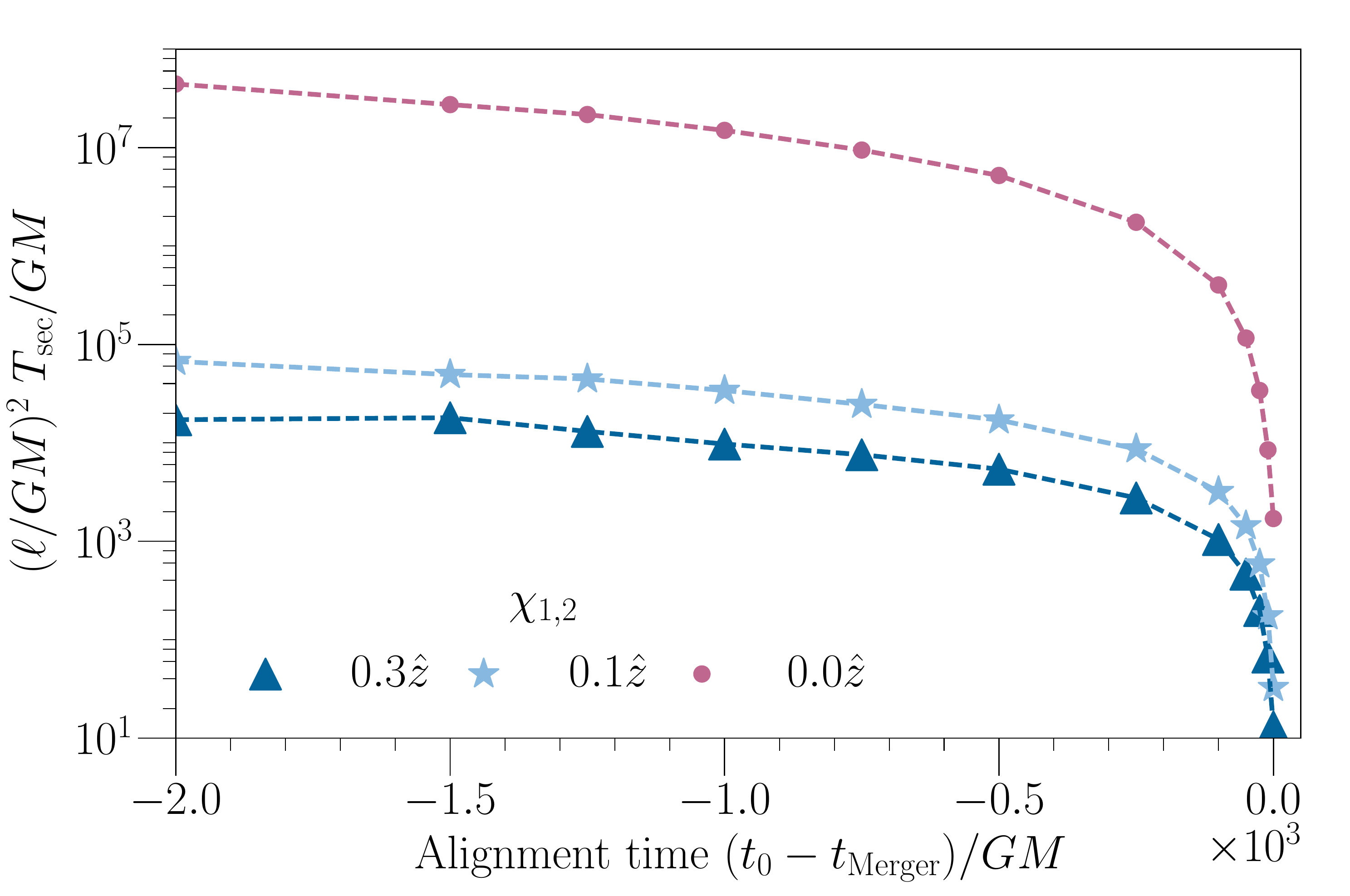}
  \caption{%
    Estimate of secular regime of validity from dephasing time
    $T_{\mathrm{sec}}$, Eq.~\eqref{eq:T-sec-def}.
    The perturbative scheme is valid within a sufficiently short time
    window $|t-t_{0}| \ll T_{\mathrm{sec}}$ about an alignment time
    $t_{0}$.  For longer times, multiple-scale analysis or
    renormalization will be needed to extend the regime of validity.
    The dephasing time is parametrically longer than the GR radiation
    reaction time,
    $T_{\mathrm{sec}} \sim T^{\textrm{GR}}_{\mathrm{RR}} (\ell/GM)^{-2} v^{-2}$.
    As expected it shrinks toward merger, remaining nonzero.
  }
  \label{fig:DephasingTime}
\end{figure}
}

\newcommand{\figDephasing}{%
\begin{figure}
  \centering
  \includegraphics[width=\columnwidth]{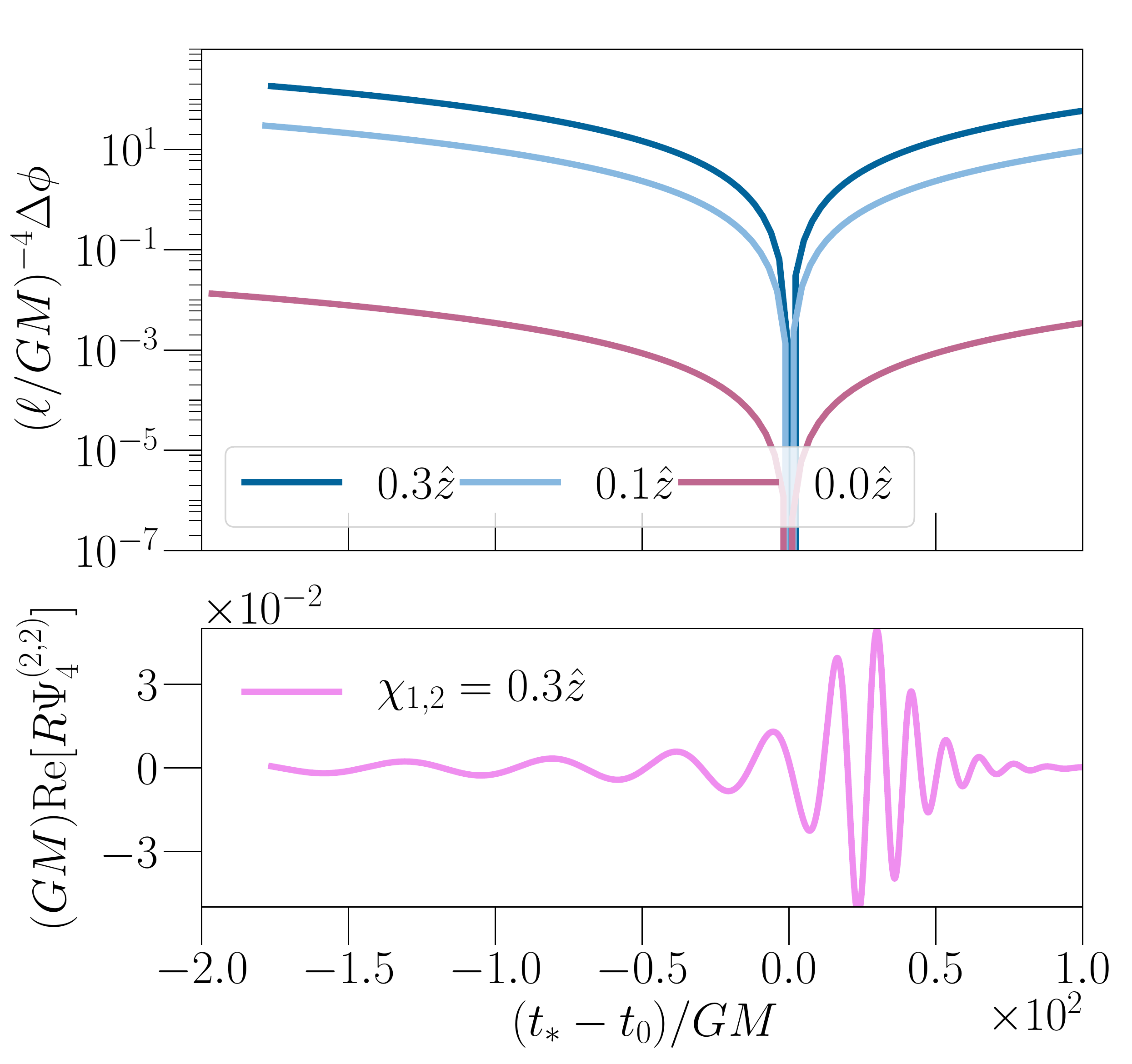}
  \caption{%
    Estimated orbital phase difference (top) for the three different
    simulations as a function of time, given by the quadratic approximation
    Eq.~\eqref{eq:Delta-phi-approx}.  We choose the alignment time
    $t_{0}$ to be when the common apparent horizon forms, the last
    time when we have access to the orbital frequency.
    From $\Delta\phi$ we can estimate how large $\ell$ must be for a
    detectable deviation from GR, or project bounds on $\ell$ for
    GR-consistent detections.
    For reference, we also plot the gravitational waveform (bottom)
    from the spin 0.3 simulation, with approximately 5 cycles of inspiral
    before merger.  This is approximately how many cycles were seen
    in GW150914~\cite{Abbott:2016blz}.  The two other simulations'
    gravitational waveforms are similar.
    }
  \label{fig:Dephasing}
\end{figure}
}

\newcommand{\figConstraints}{%
\begin{figure}
  \centering
  \includegraphics[width=\columnwidth]{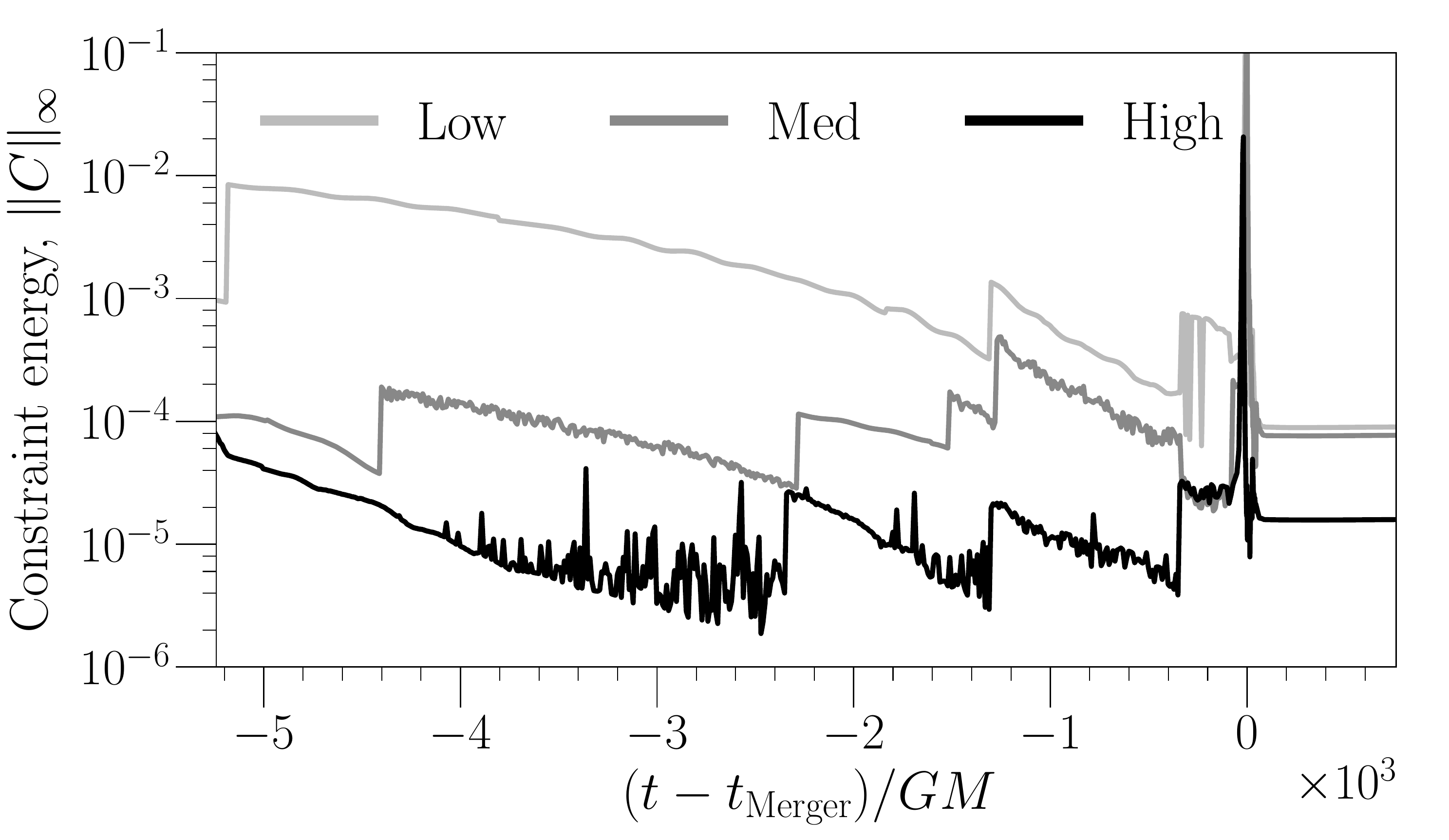}
  \caption{%
    Numerical error convergence for the highest spin (0.3
    $\hat{z}$) simulation performed in this study, which shows the greatest
    level of constraint violation.  We plot the $L^\infty$ norm of the
    constraint energy defined in Eq.~\eqref{eq:KGCE} for the low,
    medium and high numerical resolutions (adding a
    constant number of angular and radial basis functions to increase
    resolution). 
    Note that these constraints are not normalized,
    but the relative error between the levels shows exponential
    convergence. The constraint energy increases at merger, which also
    happens in the metric evolution system, and is consistent with
    other BBH simulations.
  }
  \label{fig:0p3_0p3_Constraint}
\end{figure}
}

\begin{document}

\title{Numerical binary black hole mergers in dynamical Chern-Simons:\\
  I.~Scalar field}

\author{Maria Okounkova}
\Caltech
\author{Leo C.~Stein}
\Caltech
\author{Mark A.~Scheel}
\Caltech
\author{Daniel A.~Hemberger}
\Caltech

\hypersetup{pdfauthor={Okounkova et al.}}

\date{\today}

\begin{abstract}
Testing general relativity in the non-linear, dynamical, strong-field
regime of gravity is one of the major goals of gravitational wave
astrophysics.
Performing precision tests of general relativity (GR) requires
numerical inspiral, merger, and ringdown waveforms for binary black
hole (BBH) systems in theories beyond GR.
Currently, GR and scalar-tensor gravity are the only theories
amenable to numerical simulations.
In this article, we present a well-posed perturbation scheme for
numerically integrating beyond-GR theories that have a continuous
limit to GR.
We demonstrate this scheme by simulating BBH mergers in dynamical
Chern-Simons gravity (dCS), to linear order in the perturbation
parameter.
We present mode waveforms and energy fluxes of the dCS pseudoscalar
field from our numerical simulations.
We find good agreement with analytic predictions
at early times, including the absence of pseudoscalar dipole
radiation.
We discover new phenomenology only accessible through numerics:
a burst of dipole radiation during merger.
We also quantify the self-consistency of the perturbation
scheme.
Finally, we estimate bounds that GR-consistent LIGO detections could
place on the new dCS length scale, approximately
$\ell \lesssim \mathcal{O}(10)~\mathrm{km}$.
\end{abstract}


\maketitle

\section{Introduction}

General relativity has been observationally and experimentally tested
for almost a century, and has been found consistent with all precision
tests to date~\cite{Will:2014xja}.  But no matter how well a theory
has been tested, it may be invalidated at any time when pushed to a
new regime.  Indeed, there are many theoretical reasons to believe
that general relativity (GR) cannot be the ultimate description of
gravity, from non-renormalizability to the black hole information
problem.

Moreover, from the empirical standpoint, all \emph{precision} tests of
GR to date have been in the slow-motion, weak-curvature regime.  With
the Laser Interferometer Gravitational Wave Observatory (LIGO) now
detecting the coalescence of compact binary
systems~\cite{Abbott:2016blz, Abbott:2016nmj, Abbott:2017vtc}, we finally have direct
access to the non-linear, dynamical, strong-field regime of gravity.
This is an arena where GR lacks precision tests, and it may give clues
to a theory beyond GR.  The LIGO collaboration has already used the
detections of GW150914, GW151226, and GW170104 to perform some tests of
GR~\cite{TheLIGOScientific:2016src, Abbott:2017vtc}, but these are not yet very
precise: a model-independent test gives 96\% agreement with GR.

Both black hole (BH) and neutron star (NS) binaries probe the
strong-field regime.  However, NSs have the added complication that
the equation of state of dense nuclear matter is presently unknown.
Until more is known about the equation of state, we must rely on
binary black holes (BBHs) for precision tests of GR.
Yunes, Yagi, and Pretorius argued~\cite{Yunes:2016jcc} that the lack
of understanding of BBH merger in beyond-GR theories severely limits
the ability to constrain gravitational physics using GW150914 and
GW151226.
Thus, to perform tests
of GR with BBHs, we require inspiral, merger, and ringdown waveform
predictions for these systems, which can only come from numerical
simulations.

To date, BBH simulations have only been performed in GR and
scalar-tensor gravity~\cite{Berti:2015itd} (note that BBHs in massless
scalar-tensor gravity will be identical to GR, under ordinary initial
and boundary conditions).  There are a huge number of beyond-GR
theories~\cite{Berti:2015itd}, and for the vast majority of them,
there is no knowledge of whether there is a well-posed initial value
formulation, a necessity for numerical simulations.  Indeed, there is
evidence that dynamical Chern-Simons gravity, the
beyond-GR theory we use here as an example, lacks a well-posed initial
value formulation~\cite{Delsate:2014hba}.

Our goal is to numerically integrate BBH inspiral, merger,
and ringdown in theories beyond GR that are viable but that do not
necessarily have a well-posed initial value problem.  This goal is relevant even for those
only interested in parametric, model-independent tests, because there
is presently no theory guidance for late-inspiral and merger waveforms
in theories beyond GR.

We are only interested in theories that are sufficiently ``close'' to
GR: for a theory to be viable, it has to be able to pass all the tests
that GR has passed.  This motivates an effective field-theory (EFT)
approach.  We assume that there is a high-energy theory whose
low-energy limit gives GR plus ``small'' corrections.  The effective
theory of GR with corrections does not need to capture arbitrarily
short-distance physics.  Such a theory is valid up to some cutoff, and
modes shorter than this distance scale are said to be outside of the
regime of validity of the EFT.  The EFT only needs to be well-posed
for the modes within the regime of validity.  This can be accomplished
with perturbation theory.

We present a perturbation scheme for numerically integrating beyond-GR
theories that limit to GR.  For such a theory, we perturb it about GR
in powers of the small coupling parameter.  We collect equations of
motion at each order in the coupling, creating a tower of equations,
with each level inheriting the same principal part as the background
GR system.  The well-posedness of the initial value problem in
GR~\cite{Wald} thus ensures the well-posedness of this framework, even
if the ``full'' underlying theory may not have a well-posed initial
value formulation.

In this study, we apply our perturbation framework to BBH mergers in
dynamical Chern-Simons gravity (dCS)~\cite{Alexander:2009tp}, to
linear order in perturbation theory.  This theory involves a
pseudoscalar field coupled to the parity-odd Pontryagin
curvature invariant with a small coupling parameter, and at linear
order gives a scalar field evolving on a GR BBH background.

There are a number of theoretical motivations for considering
dynamical Chern-Simons.  The dCS interaction arises when cancelling
gravitational anomalies in chiral theories in curved
spacetime~\cite{Delbourgo:1972xb, Eguchi:1976db, AlvarezGaume:1983ig},
including the famous Green-Schwarz anomaly cancellation in string
theory~\cite{Green:1984sg} when compactified to four
dimensions~\cite{Alexander:2009tp, polchinski1998string1,
  polchinski1998string2}.  DCS also arises in loop quantum gravity
when the Barbero-Immirzi parameter is allowed to be a spacetime
field~\cite{Taveras:2008yf, Mercuri:2009zt}.  From an EFT standpoint, dCS is the
lowest-mass-dimension correction that has a parity-odd interaction.
All other EFTs at the same mass dimension have parity-even
interactions, so the phenomenology of dCS is
distinct~\cite{Yagi:2011xp}.
The dCS interaction was also included in Weinberg's EFT of
inflation~\cite{Weinberg:2008hq}.

From a practical standpoint, there are already a large number of dCS
results in the literature that we can compare
against~\cite{Yunes:2009hc, Yagi:2011xp, Yagi:2012ya, Yagi:2012vf,
  Yagi:2013mbt, Konno:2014qua, Stein:2014xba}, including post-Newtonian (PN)
calculations for the BBH inspiral.  One of the more
important results is that scalar dipole radiation is highly suppressed
in dCS during the inspiral~\cite{Yagi:2011xp}.  Dipole radiation is present in
scalar-tensor theory and Einstein-dilaton-Gauss-Bonnet (EdGB), and
enters with two fewer powers of the orbital velocity (i.e.~1
PN order earlier) than the leading quadrupole radiation of
GR.  This leads to gross modifications of the inspiral, but dCS avoids
this problem because the dipole is suppressed.  As a result, the perturbative treatment of dCS will be
valid for a longer period of inspiral than scalar-tensor or EdGB.

The paper is organized as follows.
Section~\ref{sec:formalism} covers the analytical and numerical
formalisms.  More specifically, in Sec.~\ref{sec:intro-chern-simons}
we introduce dynamical Chern-Simons, and in
Sec.~\ref{sec:order_reduction} we present the perturbation scheme, 
which is valid for any theory with a continuous limit to GR.
We discuss the numerical scheme in Sec.~\ref{sec:numerical_scheme}
(some numerical details are in the Appendix).
We present the results of numerically implementing this formalism
in dCS on
three different binary mergers in Sec.~\ref{sec:results}.
Section~\ref{sec:Phenomenology} reviews some previously-known analytic
phenomenology of the BBH inspiral problem in dCS.
Section~\ref{sec:Waveforms} presents the waveform results,
and~\ref{sec:EnergyFluxes} presents the energy fluxes, both including
comparison to PN.
In Sec.~\ref{sec:RegimeOfValidity} we use the numerical results
to assess the validity of the perturbation scheme.
In Sec.~\ref{sec:Detectability} we use the numerical results to
estimate the detectability of dCS and the bounds that could be placed
by LIGO detections.
We conclude and discuss in Sec.~\ref{sec:discussion}, and lay out
plans for future work.

\section{Formalism}
\label{sec:formalism}

Throughout this paper, we set $c = 1$ and $\hbar=1$ so that
$[M]=[L]^{-1}$.
Since there will be more than one length scale, we explicitly include
factors of the reduced Planck mass $\mpl^{-2} = 8\pi G$ and the ``bare''
gravitational length $GM$, though quantities in our code are
non-dimensionalized with $GM=1$.
Latin letters in the middle of alphabet $\{i,j,k,l,m,n\}$ are
(3-dimensional) spatial indices, while Latin letters in the beginning
of the alphabet $\{a,b,c,d\}$ refer to (4-dimensional) spacetime
indices.
We follow the sign conventions of~\cite{Wald}, and $g_{ab}$ refers to
the 4-dimensional spacetime metric, with signature
\mbox{$(-{}+{}+{}+)$}, and with $\cd$ its Levi-Civita connection.

\subsection{Action and equations of motion}
\label{sec:intro-chern-simons}

The method we present in this paper applies to a large
number of beyond-GR theories that have a continuous limit to GR, 
but for concreteness we focus on dCS.  We start with the
four-dimensional action
\begin{align}
  \label{eq:action-0}
  I = \int d^{4}x \sqrt{-g}
  \left[
  L_{\txt{EH}} + L_{\vartheta} + L_{\INT} + L_{\txt{mat}} + \ldots
  \right]
  \,,
\end{align}
where the omitted terms $(\ldots)$ are above the cutoff of our
EFT treatment.  Here $g$ without indices is
the determinant of the metric, $L_{\txt{EH}}$ is the Einstein-Hilbert
Lagrangian, $L_{\vartheta}$ is the Lagrangian of a minimally coupled
(pseudo-{})scalar field $\vartheta$ (also referred to in the literature
as the axion), 
$L_{\INT}$ is a beyond-GR
interaction between $\vartheta$ and curvature terms, and
$L_{\txt{mat}}$ is the Lagrangian for ordinary matter.  In this
paper, we are considering a binary black hole (BBH) merger in
dCS, so we ignore $L_{\txt{mat}}$.

Explicitly, these action terms are given by
\begin{subequations}
  \begin{align}
    L_{\txt{EH}} &= \tfrac{\mpl^{2}}{2} R \,,
    &
    L_{\vartheta} &= -\tfrac{1}{2} (\pd\vartheta)^{2} \,,
    \\
    L_{\INT} &= -\tfrac{\mpl}{8} \ell^{2} \vartheta \ \pont
               \,.
  \end{align}
\end{subequations}
Here the Ricci scalar of $g_{ab}$ is $R$.
With our unit system, $[g]=[L]^{0}$, coordinates carry dimensions
of length, $[x]=[L]^{1}$, and note that the scalar field
$\vartheta$ has
been canonically normalized, $[\vartheta] = [L]^{-1}$.  We have
omitted any potential $V(\vartheta)$, so $\vartheta$ is massless and
long-ranged, as appropriate for a ``gravitational'' degree of
freedom. In the interaction Lagrangian $L_{\INT}$, the scalar field
$\vartheta$ is coupled to the 4-dimensional Pontryagin
density (also known as the Chern-Pontryagin density) $\pont$,
\begin{align}
  \label{eq:pont}
  \pont \equiv {}^{*} R^{abcd} R_{abcd}
  = \tfrac{1}{2} \epsilon^{abef} R_{ef}{}^{cd} R_{abcd}
  \,,
\end{align}
where $\epsilon_{abcd}$ is the fully antisymmetric Levi-Civita
tensor.

The coupling strength of this interaction is governed by the new
parameter $\ell$ with dimensions of length.  This parameter takes on
specific values if this EFT arises from the low-energy limit of
certain string theories~\cite{Green:1984sg} or to cancel gravitational
anomalies~\cite{AlvarezGaume:1983ig, polchinski1998string1, polchinski1998string2}.  However, here we simply take it as a ``small''
coupling parameter.  In the limit that $\ell\to 0$, we recover general
relativity with a massless, minimally coupled scalar field.

The coupling parameter conventions vary throughout the literature.
To enable comparisons, we express the couplings of a number of works
in terms of our conventions.  To put Yagi et al.~\cite{Yagi:2011xp} into
our conventions, use
\begin{align}
  \label{eq:YagiConversion}
  \kappa^{\txt{YSYT}} &= \frac{1}{2} \mpl^{2} \,, &
  \alpha_4^\txt{YSYT} &= -\frac{\mpl \ell^2}{8} \,, &
  \beta^{\txt{YSYT}} &=1 \,.
\end{align}
To convert Alexander and Yunes~\cite{Alexander:2009tp} into our
conventions,
\begin{align}
  \label{eq:AYConversion}
  \kappa^{\txt{AY}} &= \frac{1}{2} \mpl^{2} \,, &
  \alpha_4^\txt{AY} &= +\frac{\mpl \ell^2}{2} \,, &
  \beta^{\txt{AY}} &=1 \,.
\end{align}
To compare with McNees et al.~\cite{McNees:2015srl}, use
\begin{align}
  \label{eq:MSYConversion}
  \kappa^{\txt{MSY}} &= \mpl^{-1} \,, &
  \alpha^\txt{MSY} &= +\frac{\ell^2}{2} \,.
\end{align}
The conventions of Stein~\cite{Stein:2014xba} agree with ours (except
for an inconsequential sign change in the definition of $\pont$, which
is compensated for by an additional sign everywhere $\pont$ appears).

Below we will perform an expansion in powers of $\ell^{2}$.  To
simplify matters, we insert a dimensionless formal order-counting
parameter $\eps$ that will keep track of powers of $\ell^{2}$.
Expanding in a dimensionless parameter ensures that field quantities
at different orders have the same length dimension.

Specifically, we replace the action in Eq.~\eqref{eq:action-0}
with
\begin{align}
  \label{eq:action-with-eps}
  I_{\eps} = \int d^{4}x \sqrt{-g}
  \left[
  L_{\txt{EH}} + L_{\vartheta} + \eps L_{\INT} + L_{\txt{mat}} + \ldots
  \right]
  \,,
\end{align}
a one-parameter family of actions parameterized by $\eps$.  Formally,
we recover the action in Eq.~\eqref{eq:action-0} when $\eps=1$.

Varying the action Eq.~\eqref{eq:action-with-eps} with respect to the
scalar field, we have the sourced wave equation
\begin{align}
\square \vartheta =  \eps \tfrac{\mpl}{8} \ell^2 \ \pont \,,
\label{eq:eom-full-scalar}
\end{align}
where $\square = \nabla_a \nabla^a$ is the d'Alembertian operator.
Varying with respect to the metric gives the corrected Einstein field
equations,
\begin{align}
\label{eq:eom-full-metric}
\mpl^2 G_{ab} + \mpl \eps \ell^2 C_{ab} = T^{\vartheta}_{ab} + T^{\txt{mat}}_{ab}
\,,
\end{align}
where $G_{ab}$ is the Einstein tensor of $g_{ab}$, and the tensor
$C_{ab}$ includes first and second derivatives of $\vartheta$, and
second and \emph{third} derivatives of the metric,
\begin{align}
C_{ab} \equiv \epsilon_{cde(a}\nabla^{d} R_{b)}{}^{c} \nabla^e \vartheta + {}^*R^c{}_{(ab)}{}^d \nabla_c \nabla_d \vartheta.
\label{eq:CTensor}
\end{align}
Since we are focusing on BBH mergers, $T^{\txt{mat}}_{ab}=0$.
The scalar field's stress-energy tensor $T^{\vartheta}_{ab}$ is given by
the expression for a canonical, massless Klein-Gordon field,
\begin{align}
T^{\vartheta}_{ab} = \nabla_a \vartheta \nabla_b \vartheta
  - \frac{1}{2} g_{a b} \nabla_c \vartheta \nabla^c \vartheta
\,.
\label{eq:StressEnergy}
\end{align}
From here forward we will drop the superscript $\vartheta$.

The ``full'' system of equations for dCS is thus the pair of
Eqs.~\eqref{eq:eom-full-scalar} and \eqref{eq:eom-full-metric}.

\subsection{Perturbation scheme}
\label{sec:order_reduction}

Because $C_{ab}$ in Eq.~\eqref{eq:eom-full-metric} contains
third derivatives of the metric, the ``full'' system of
equations for dCS likely lacks a well-posed initial value
formulation~\cite{Delsate:2014hba}.  In the language of particle
physics, this is equivalent to the appearance of ghost modes above a
certain momentum scale~\cite{Dyda:2012rj}.

From the EFT point of view, though, the ghost modes and ill-posedness
are nothing more than the
breakdown of the regime of validity of the theory, which should be
valid for long wavelength modes in the decoupling limit $\ell\to 0$.
To excise the ghost modes and arrive at a well-posed initial value
formulation, we expand about $\eps=0$, which is simply GR coupled
to a massless minimally-coupled scalar field and certainly has a
well-posed initial value problem~\cite{Wald}.  As a result,
all higher orders in $\eps$ will inherit the well-posedness of the
zeroth-order theory, by inheriting the principal parts of the
differential equations.

We begin this order-reduction scheme by
expanding the metric and scalar field in power series in $\eps$,\footnote{%
Note that this is not a Taylor series, since there is no factor of
$1/k!$ in the $k$th term.  These factors must be tracked if using
standard perturbation theory, e.g.~with the \codename{xPert}
package~\cite{xActWebPage, Brizuela:2008ra}.
}
\begin{subequations}
  \begin{align}
    g_{ab} &= g_{ab}^\mathrm{(0)} + \sum_{k = 1}^{\infty}
             \eps^k h_{ab}^{(k)}
             \,, \label{eq:MetricExpansion} \\
    \vartheta &= \sum_{k=0}^{\infty} \eps^k \vartheta^{(k)}
                \,.
                \label{eq:ScalarFieldExpansion}
  \end{align}
\end{subequations}
Note that since $\eps$ is dimensionless, each $\vartheta^{(k)}$ has
the same units as $\vartheta$, and similarly for $h_{ab}^{(k)}$.  This
expansion is now inserted into the field equations, which are likewise
expanded in powers of $\eps$, and we collect orders homogeneous in
$\eps^{k}$, as below.  This results in a ``tower'' of systems of
equations that must be solved at progressively increasing orders in
$\eps$.  This scheme is quite general, and should apply to any theory
that has a continuous limit to GR.

\subsubsection{Order $\eps^{0}$}
Zeroth order comes from taking $\eps\to 0$, which simply gives the
system of GR coupled to a massless, minimally coupled scalar field,
\begin{subequations}
  \label{eq:order-0-sys}
  \begin{align}
    \label{eq:eom-met-0}
    \mpl^{2} G_{ab}[g^{(0)}] &= T^{(0)}_{ab} \,, \\
    \label{eq:eom-vartheta-0}
    \square^{(0)} \vartheta^{(0)} &= 0 \,,
  \end{align}
\end{subequations}
where $G_{ab}[g^{(0)}]$ is the Einstein tensor of the background
metric $g^{(0)}$, $\square^{(0)}$ is the associated d'Alembert
operator, and $T^{(0)}$ is the stress-energy of
$\vartheta^{(0)}$.  This system certainly has a well-posed initial
value problem.

Because of the explicit presence of $\eps$ in front of $L_{\INT}$ in
the action [Eq.~\eqref{eq:action-with-eps}], $C_{ab}$ does not appear in
the metric equation~\eqref{eq:eom-met-0}, and the
Pontryagin source does not appear on the right-hand side of the scalar
equation~\eqref{eq:eom-vartheta-0}.  These terms have been pushed to
one order higher and will appear below.

On general grounds, we expect that any initially non-vanishing scalar
field will radiate away within a few dynamical times.  Similarly, if
we start with a $\vartheta^{(0)}= 0$ initial condition and
impose purely outgoing boundary conditions, $\vartheta^{(0)}$ will
remain zero throughout the entire simulation.  Therefore, rather than
simulating a vanishingly small $\vartheta^{(0)}$, we simply
analytically assume that $\vartheta^{(0)} = 0$.

Therefore, at order $\mathcal{O}(\eps^{0})$, the system will simply be
\begin{align}
  \label{eq:order-0-g-eq}
  G_{ab}[g^{(0)}] = 0 \,,
\end{align}
and the solution will be
\begin{align}
  \label{eq:bg-fields}
  (g^{(0)},\vartheta^{(0)}) = (g^{\GR}, 0) \,,
\end{align}
where $g^{\GR}$ is a GR solution to the BBH inspiral-merger-ringdown
problem.

\subsubsection{Order $\eps^{1}$}
Continuing to linear order in $\eps$, we find the system
\begin{subequations}
  \label{eq:order-1-sys-general}
  \begin{align}
    \mpl^{2} G^{(1)}_{ab}[h^{(1)}; g^{(0)}] &=
    - \mpl \ell^{2} C_{ab}^{(0)} + T_{ab}^{(1)}
    \,, \\
    \square^{(0)} \vartheta^{(1)} + \square^{(1)} \vartheta^{(0)} &=
    \tfrac{\mpl}{8} \ell^{2} [ \pont ]^{(0)}
    \,.
  \end{align}
\end{subequations}
As noted above, the explicit presence of $\eps$ in the
action~\eqref{eq:action-with-eps} and equations of motion
[\eqref{eq:eom-full-scalar} and \eqref{eq:eom-full-metric}] lead to
$C^{(0)}$ and $[\pont]^{(0)}$ appearing in these $\eps^{1}$
equations strictly as source terms.  By construction, the principal
part of this differential system is the same as the principal part of
the $\mathcal{O}(\eps^{0})$ system, and thus it inherits its
well-posedness property.  This is true at all higher orders in
perturbation theory.

Here, $G^{(1)}[h^{(1)}; g^{(0)}]$ is the linearized
Einstein operator, built with the covariant derivative $\cd^{(0)}$
compatible with $g^{(0)}$, acting on the metric deformation $h^{(1)}$.
The d'Alembert operator receives the correction $\square^{(1)}$, which
depends on the metric deformation $h^{(1)}$.
The quantity $C^{(0)}_{ab}$ is the same as the definition given in
Eq.~\eqref{eq:CTensor}, evaluated on the background quantities
$(g^{(0)},\vartheta^{(0)})$.  Similarly, $[\pont]^{(0)}$ is the
Pontryagin density evaluated on the background spacetime metric
$g^{(0)}$.  Finally, $T^{(1)}_{ab}$ is the first-order perturbation to
the stress-energy tensor; since $T_{ab}$ is quadratic in $\vartheta$,
$T^{(1)}_{ab}$ has pieces both linear and quadratic in
$\vartheta^{(0)}$ (the quadratic-in-$\vartheta^{(0)}$ pieces are
linear in $h^{(1)}$).

The crucial property at this order is that both $C^{(0)}$ and
$T^{(1)}$ are built from pieces linear and quadratic in
$\vartheta^{(0)}$.  At order $\mathcal{O}(\eps^{0})$, we found that
$\vartheta^{(0)}=0$.  Therefore, when evaluated on the
$\mathcal{O}(\eps^{0})$ solution [Eq.~\eqref{eq:bg-fields}], these
both vanish,
\begin{align}
  C^{(0)}_{ab}[\vartheta^{(0)}=0] &= 0 \,, &
  T^{(1)}_{ab}[\vartheta^{(0)}=0] &= 0 \,.
\end{align}
Therefore, at order $\mathcal{O}(\eps^{1})$ in perturbation theory,
evaluating on the background solution, we have the system
\begin{subequations}
  \label{eq:order-1-sys}
  \begin{align}
    \label{eq:order-1-h-eq}
    \mpl^{2} G^{(1)}_{ab}[h^{(1)}; g^{(0)}] &= 0
    \,, \\
    \label{eq:order-1-vj-eq}
    \square^{(0)} \vartheta^{(1)} &=
    \tfrac{\mpl}{8} \ell^{2} [ \pont ]^{(0)}
    \,.
  \end{align}
\end{subequations}

In the metric perturbation equation~\eqref{eq:order-1-h-eq}, starting
with $h^{(1)} = 0$ initial conditions and imposing purely outgoing boundary
conditions will enforce $h^{(1)}=0$ throughout the entire simulation.
Similarly, we can argue that small perturbations of $h^{(1)}$ would
radiate away on a few dynamical times, since there is no potential to
confine the metric perturbations.  Once again, rather than simulating
a vanishingly small field, we will just analytically assume that
$h^{(1)}=0$.  Therefore, at order $\mathcal{O}(\eps^{1})$, there is no
metric deformation, and the system is only
Eq.~\eqref{eq:order-1-vj-eq}, driven by the background
system~\eqref{eq:order-0-g-eq} which generates the source term
$[\pont]^{(0)}$.

\subsubsection{Order $\eps^{2}$}
\label{sec:ordereps2}
This perturbation scheme can be extended to any order desired.
Although this paper reports only on work extending through
$\mathcal{O}(\eps^{1})$, we sketch the derivation of
$\mathcal{O}(\eps^{2})$, since that is the lowest order where a metric
deformation is sourced.

Schematically, the system at $\mathcal{O}(\eps^{2})$, after accounting
for the vanishing of $\vartheta^{(0)}$ and $h^{(1)}$, is
\begin{subequations}
  \label{eq:order-2-sys-sketch}
  \begin{align}
    \label{eq:order-2-sys-h}
    \mpl^{2}G^{(1)}_{ab}[h^{(2)}] &=
    - \mpl \ell^{2} C_{ab}^{(1)}[\vartheta^{(1)}] +
    T_{ab}^{(2)}[\vartheta^{(1)}, \vartheta^{(1)}] \,, \\
    \label{eq:order-2-sys-vj}
    \square^{(0)} \vartheta^{(2)} &= 0
    \,.
  \end{align}
\end{subequations}
The operator $C^{(1)}[\vartheta^{(1)}]$ is linear in its argument, and
$T^{(2)}[\vartheta^{(1)},\vartheta^{(1)}]$ is linear in each slot.
Various other combinations have vanished.
In~\eqref{eq:order-2-sys-h}, vanishing source terms were quadratic in
$h^{(1)}$ or built from the product of
$h^{(1)} \times \vartheta^{(1)}$.  In~\eqref{eq:order-2-sys-vj},
$\ell^{2} [\pont]^{(1)}$ is proportional to $h^{(1)}$ and thus
vanishes, as do terms such as $\square^{(1)} \vartheta^{(1)}$ (linear
in $h^{(1)}$) and $\square^{(2)} \vartheta^{(0)}$ (linear in
$\vartheta^{(0)}$).

We leave detailed discussion of order $\mathcal{O}(\eps^{2})$ to future
work~\cite{OkounkovaSteinForthcoming}.

\subsubsection{Summary and scaling}
\label{sec:summary-scaling}

Let us briefly summarize the perturbative order-reduction scheme and
discuss the scaling of different orders.  The system at orders
$\eps^{0}$ and $\eps^{1}$ is
\begin{subequations}
\label{eq:summary-system}
  \begin{align}
    \label{eq:summary-system-0}
    \mathcal{O}(\eps^{0}): &&
    G_{ab}[g^{(0)}] &= 0 \,, &
    \vartheta^{(0)} &= 0 \,, \\
    \label{eq:summary-system-1}
    \mathcal{O}(\eps^{1}): &&
    \square^{(0)} \vartheta^{(1)} &= \tfrac{\mpl}{8} \ell^{2} [\pont]^{(0)} \,, &
    h^{(1)} &= 0 \,, \\
\intertext{and if we were to continue to $\mathcal{O}(\eps^{2})$,}
    \label{eq:summary-system-2}
    \mathcal{O}(\eps^{2}): &&
    G^{(1)}_{ab}[h^{(2)}] &= \mpl^{-2} T^{\txt{eff}}_{ab}\,, &
    \vartheta^{(2)} &= 0 \,,
  \end{align}
\end{subequations}
where $T^{\txt{eff}}_{ab}$ may be determined from the right hand side of
Eq.~\eqref{eq:order-2-sys-h}.

Zeroth order \eqref{eq:summary-system-0} is just vacuum GR, which has
no intrinsic scale.  As is very common in numerical relativity
simulations, the coordinates used in the simulation are dimensionless 
and in units of
the total ADM mass, $X^{a} = x^{a}/(GM)$.  This means
that $\cd$ may be non-dimensionalized by pulling out a factor of
$(GM)^{-1}$, Riemann may be non-dimensionalized by pulling out a
factor of $(GM)^{-2}$, etc.

Meanwhile, the new length scale and coupling parameter $\ell$ enters
at first order.  If we non-dimensionalize the derivative operator and
curvature tensors in Eq.~\eqref{eq:summary-system-1}, we will find
\begin{align}
  (GM)^{-2} \square^{(0)} \vartheta^{(1)} =
  \frac{\mpl}{8} \ell^{2} (GM)^{-4} [\pont]^{(0)} \,.
\end{align}
We therefore define the dimensionless scalar field $\Psi$ via
\begin{align}
  \label{eq:PsiConversion}
  \vartheta^{(1)} = \frac{\mpl}{8}
  \left( \frac{\ell}{GM} \right)^{2} \Psi
  \,.
\end{align}
Then $\Psi$ will satisfy
\begin{align}
  \label{eq:UnitsScalarFieldEOM}
  \square^{(0)} \Psi = [\pont]^{(0)} \,.
\end{align}
Thus the analytic dependence of $\vartheta^{(1)}$ on $(\ell/GM)$ has
been extracted.  The solution $\Psi$ can later be scaled
to reconstruct $\vartheta^{(1)}$ for any allowable value of
$(\ell/GM)$.

All of the results that we present will be given in terms of powers of
the dimensionless coupling $(\ell/GM)$.  We will also compare to known
post-Newtonian results~\cite{Yagi:2013mbt}, that were presented in
terms of $\alpha_{4}^{\txt{YSYT}}$.  To perform the comparison, we use
the conversion given in Eq.~\eqref{eq:YagiConversion}.

Finally, though we do not address $\mathcal{O}(\eps^{2})$ simulations
in this paper, we should still study how $h^{(2)}$ scales with
$\ell$ and $(GM)$.  Since the perturbative scheme preserves the units
of length of fields, $[h^{(k)}] = [g] = [L]^{0}$ is already
dimensionless; however, it still depends on $(\ell/GM)$ in a specific
way.  When we move to  units in which we measure lengths and times in
units of $(GM)$, we find it is appropriate to define a scaled metric
deformation $\Upsilon$ via
\begin{align}
  \label{eq:Upsilon-def}
  h^{(2)}_{ab} \equiv \left( \frac{\ell}{GM} \right)^{4} \Upsilon_{ab} \,.
\end{align}
Then this dimensionless quantity $\Upsilon$ will satisfy an
equation that is schematically
\begin{align}
\label{eq:Upsilon-EOM}
  \cd^{2} \Upsilon + \mathrm{L.O.T.}
  \sim (\cd\Psi)^{2} + (\cd\Psi)(\cd R) + (\cd^{2}\Psi) R \,,
\end{align}
where L.O.T.~stands for lower order terms, and all derivatives and
curvatures are $\mathcal{O}(\eps^{0})$ dimensionless quantities.

\subsection{Numerical scheme}
\label{sec:numerical_scheme}

For the order $\eps^1$ part of the order reduction scheme, our overall
goal is to solve Eq.~\eqref{eq:UnitsScalarFieldEOM} on a dynamical background
metric. We co-evolve the metric and the scalar field, where
Eq.~\eqref{eq:UnitsScalarFieldEOM} is driven by Eq.~\eqref{eq:summary-system-0}.
The whole system is simulated using the Spectral Einstein Code
(\codename{SpEC})~\cite{SpECwebsite},
which uses the generalized harmonic formulation of general relativity in a
first-order, constraint-damping system~\cite{Lindblom2006} in order to
ensure well-posedness and hence numerical stability. We have added
a scalar field module that is similarly a first-order, constraint-damping system,
following~\cite{Holst:2004wt}, as outlined in App.~\ref{appendix:ScalarField}.

The code uses pseudospectral methods on an adaptively-refined grid~\cite{Lovelace:2010ne, Szilagyi:2014fna},
and thus numerical convergence with resolution of both the metric variables and the scalar field is 
exponential. We demonstrate the numerical convergence of the scalar field in App.~\ref{appendix:ScalarField}. 

The initial data for the binary black hole background is a
superposition of two Kerr-Schild black holes with a Gaussian roll-off
of the conformal factor around each black hole~\cite{Lovelace2008}.
The initial data for the scalar field is similarly given by a
superposition of approximate dCS solutions around isolated black
holes, and is given in more detail in Sec.~\ref{sec:Waveforms}.

The metric equations are evolved in a damped harmonic gauge~\cite{Szilagyi:2009qz, Lindblom2009c},
with excision boundaries just inside the apparent horizons~\cite{Hemberger:2012jz, Scheel2014}, and
minimally-reflective, constraint-preserving boundary conditions on the outer boundary~\cite{Rinne2007}.
The scalar field system, meanwhile, uses purely outgoing boundary conditions modified to
reduce the influx of constraint violations into the computational domain~\cite{Holst:2004wt}.

The Pontryagin density source term $\pont$ is computed throughout the
simulation in a 3+1 split from the available spatial quantities as
outlined in App.~\ref{appendix:EBDecomposition}.

\section{Results}
\label{sec:results}

\subsection{Background: Phenomenology of binary\break{} black hole inspirals in dCS}
\label{sec:Phenomenology}

To give the proper context for our numerical results, we first review
the previously-known phenomenology relevant to this problem.
Analytical and numerical results are known for isolated black holes in
the decoupling limit, and analytical results are known
for the binary black hole problem in the decoupling limit and at slow
velocities ($v/c \ll 1$).

Any spherically-symmetric metric will have vanishing Pontryagin
density.\footnote{%
  This is straightforward to verify with a computer algebra system,
  using the canonical form for a spherically symmetric metric,
  $ds^{2} = - e^{2\alpha(t,r)}dt^{2} + e^{2\beta(t,r)}dr^{2} +
  r^{2}d\Omega^{2}$.
  Since it is true in this coordinate system, it is true in general.
  This is also proven in App.~A of~\cite{Grumiller:2002nm} following a
  tensorial approach.  Finally, one can appeal to a symmetry argument.
  If the metric is invariant under an $O(3)$ isometry, then the
  curvature tensor and $\pont$, being tensorial objects built only
  from $g$, must also be invariant under this symmetry.  Therefore
  $\pont$ must be a constant on each 2-sphere.  The group $O(3)$ also
  contains the reflection symmetry, sending
  points to their antipodes.  The metric is invariant under this
  reflection, but $\pont$ must flip sign, as it is a pseudo-scalar.
  But then we must have $\pont = -\pont$, so $\pont = 0$.
}
Thus the Schwarzschild solution with vanishing scalar field
is already a solution to the ``full'' dCS system.
An isolated spinning black hole in dCS,
however, is not given by the Kerr solution of
GR~\cite{Campbell:1990fu, Yunes:2009hc, Konno:2009kg, Yagi:2012ya}; the scalar field
is sourced, and the metric acquires corrections.  Analytical results
for the leading-order, small-coupling corrections to the Kerr metric
have been found in the slow-rotation approximation
($a\ll M$)~\cite{Yunes:2009hc, Konno:2009kg, Yagi:2012ya, Maselli:2017kic}.
Additionally, numerical results have been found for
the scalar field for general rotation~\cite{Konno:2014qua,
  Stein:2014wza}.  The leading-order correction
to Kerr is dipolar scalar hair, while the scalar monopole vanishes.
This vanishing scalar monopole means that
scalar dipole radiation is heavily suppressed in dCS.
At a large radius away from an isolated black hole labeled by $A$, the
dipolar scalar field goes as
\begin{align}
  \vartheta^{(1)}_{A} = \frac{\mu_{A}^{i} n_{A}^{i}}{R_{A}^{2}} \,,
\end{align}
where $R_{A}$ is the distance from black hole $A$, $n_{A}^{i}$ is the
spatial unit vector pointing away from BH $A$, and $\mu_{A}^{i}$ is
the scalar dipole moment of the BH.  This scalar dipole moment is
given by~\cite{Yagi:2011xp}
\begin{align}
\label{eq:scalar-dipole-moment}
\mu_{A}^{i} = -\frac{5}{2} \frac{\mpl \ell^2}{8} \chi_{A}^{i} \,,
\end{align}
where $\chi_{A}^{i}$ is the dimensionless spin vector of black hole $A$,
$\chi_{A}^{i} = J_{A}^{i}/G M_{A}^{2}$ (this factor of $G$ in the
denominator arises from our usage of natural units, where angular
momentum is dimensionless, $[J] = [L]^{0}$, in units of $\hbar$).

The dCS binary inspiral problem in the post-Newtonian regime ($v \ll
c$) was first treated by Yagi et al.~\cite{Yagi:2011xp}.
When two spinning BHs with scalar dipole hair are placed in proximity
with each other, the hair is responsible for a number of effects.
First, there is a correction to the binding energy due to the
dipole-dipole interaction.  Second, as the BHs orbit each other, the
net \emph{quadrupole} of the binary system has a time derivative on
the orbital timescale.  The binary's combined dipole moment is also
time-varying, but only on the spin-precession timescale, so it is
heavily suppressed.  Thus in the far zone of the binary, the scalar
field exhibits predominantly quadrupole and higher radiation, and no
$l=0$ monopole radiation.

The dominant far-zone multipole moments for the scalar field have
$|m|=l-1$ with $l\ge 2$ and the $l=1$ modes radiate on the
spin-precession timescale.  To make comparing to PN simpler, we
are simulating aligned-spin systems, so the $l=1$ mode will in fact be
non-radiative at early times.  Yagi et al.~\cite{Yagi:2011xp} gave
expressions for the scalar field $\vartheta^{(1)}$ due to spinning and
non-spinning binaries, presented in terms of symmetric tracefree (STF) tensors.
In most numerical relativity work, however, we decompose fields into
spherical harmonics,
\begin{align}
  \vartheta^{(1)\txt{FZ}} = \sum_{lm} Y_{lm}(\theta, \varphi)
  \vartheta^{(1)\txt{FZ}}_{l,m} \,.
\end{align}
Using~\cite{Blanchet:1985sp}, we convert the STF expressions
from~\cite{Yagi:2011xp,Stein:2013wza} into spherical harmonics at
extraction radius
$R$ for a \textit{spin-aligned} binary, when the post-Newtonian
approximation is valid (the early inspiral), giving
\begin{align}
\label{eq:PNExpressions}
\vartheta^{(1)\mathrm{FZ}}_{1,0} &= \sqrt{\frac{4\pi}{3}}
\frac{1}{R^2} (\mu_1 + \mu_2) \,, \\
\nonumber \vartheta^{(1)\mathrm{FZ}}_{2,1} &=
\sqrt{\frac{2\pi}{15}}
\frac{1}{R}
\left(\mu_1 \frac{m_2}{M} - \mu_2 \frac{m_1}{M}\right)
\omega (GM\omega)^{1/3} e^{-i\phi}
\,, \\
\nonumber \vartheta^{(1)\mathrm{FZ}}_{3,2} &=
\sqrt{\frac{32\pi}{105}}
\frac{1}{R}
\left(\mu_1 \frac{m_2^2}{M^2} + \mu_2 \frac{m_1^2}{M^2}\right)
\omega (GM\omega)^{2/3} i e^{-2i\phi}
\,.
\end{align}
Here $\phi = \phi(t)$ is the orbital phase, $\omega = \omega(t) =
\dot{\phi}$ is the orbital frequency, $m_A$ is the mass of each black
hole, $M = m_1 + m_2$ is the total mass,\footnote{%
  In PN literature, $m$ is often used as the total mass. We use $M$
  here in order to be consistent with numerical relativity
  literature.}
and $\mu_A$ is the $z$ component (the only component since
this calculation is for a spin-aligned binary)
of the scalar dipole moment from
Eq.~\eqref{eq:scalar-dipole-moment}.
Note that the $(1,0)$ mode is time-independent (and hence
non-radiative), since we are focusing on spin-aligned systems.

The behavior of the scalar field during the late inspiral and merger
was previously unknown, and is part of the motivation for the present
numerical study.

\subsection{Scalar field waveforms}
\label{sec:Waveforms}

\tabRuns{}

We performed three numerical simulations in this formalism, each at
low, medium, and high numerical resolutions, with parameters given by
Table~\ref{tab:runs}.  We chose three values for the BHs'
dimensionless spins of $0.0$, $0.1$, and $0.3$, to qualitatively see
the effect of spin on the physics, and to allow for comparison with
analytical calculations.  While \codename{SpEC} can simulate very high
spins~\cite{Scheel2014}, the analytics we compare against use the
small-spin expansion and stop at linear order in spin.  Therefore the
$\mathcal{O}(\chi^{2})$ errors should be at most $\sim 30$\% of the
$\mathcal{O}(\chi)$ effects we compare against.  Similarly, while
modeling spin precession is possible~\cite{OssokineEtAl:2014}, it is
not the focus of this study, and thus we have eliminated this
complication by aligning all of the spins with the orbital angular
momentum.

\figWaveformThreeThree{}
\figWaveformOneOne{}
\figWaveformZeroZero{}

As mentioned in Sec.~\ref{sec:Phenomenology}, the scalar field around
an isolated, slowly spinning black hole in dCS is approximately a
dipole.  We use this analytic approximation as the basis for our
initial data, as mentioned in Sec.~\ref{sec:numerical_scheme}.  The
initial scalar field is a superposition of two slow-rotation dipole
solutions (since all of the dimensionless spins are $\leq 0.3$), one
around each black hole.  We apply a boost to account for the initial velocity
of each black hole.  As our scalar field evolution system is
first-order (see App.~\ref{appendix:ScalarField}), we also initialize
the variables corresponding to the spatial and time derivatives of
$\Psi$ to the analytical derivatives of the approximate dipole
solution.  For the non-spinning simulation, we set the initial value
of $\Psi$ and its derivatives to zero.

We plot mode-decomposed waveforms extracted from the highest
resolution simulations of the three simulations in
Figs.~\ref{fig:0p3_0p3_Waveform}, \ref{fig:0p1_0p1_Waveform},
and~\ref{fig:0p0_0p0_Waveform}.  Each figure shows the $(l=2,m=2)$
mode of the Newman-Penrose quantity $\Psi_4$ decomposed into
spin-weight $-2$ spherical harmonics, and the dominant $(l,m=l-1)$
modes of the scalar $\vartheta^{(1)}$ for $l=1,2,3$,
along with the PN comparisons from Eq.~\eqref{eq:PNExpressions}.

We immediately see that at early times, there is good qualitative
agreement between the numerical waveforms and the PN predictions, with
the $(l=2,m=1)$ mode dominating, as expected.  In the PN formulas of
Eq.~\eqref{eq:PNExpressions}, we used the instantaneous coordinate
orbital frequency and phase calculated from the black hole trajectories 
for $\omega$ and $\phi$. Since the
starting phase is arbitrary, we perform a phase alignment (by eye)
between the numerical results and the PN waveforms.

As expected, because the spins are not precessing, there is no dipole
radiation at early times.  The offset away from zero seen in the
$(l=1,m=0)$ panel of Fig.~\ref{fig:0p3_0p3_Waveform} is a real
physical effect: it is due to the combined dipole moments of the two
individual black holes and their orbital angular momentum.  After
merger, the $l=1$ moment settles down to a new non-zero value (below
the resolution of this figure)
determined by the spin of the final black hole, again via
Eq.~\eqref{eq:scalar-dipole-moment}.  In between, there is a burst of
scalar dipole radiation.  This is a newly discovered phenomenon that could
not have been computed with analytic post-Newtonian calculations.
Scalar monopole radiation, meanwhile, is consistent with zero within
the numerical errors of the simulation.

\subsection{Energy fluxes} 
\label{sec:EnergyFluxes} 

Having solved for the scalar field $\vartheta^{(1)}$, we can evaluate
physical quantities such as its stress-energy tensor,
Eq.~\eqref{eq:StressEnergy}.  From $T_{ab}^{(\vartheta)}$, we can
compute the energy flux through some 2-sphere $S^{2}_{R}$ at
coordinate radius $R$ via
\begin{align}
\label{eq:AnalyticalEnergyFlux}
\dot{E}^{(\vartheta)} =
  \int_{S_R^2}  T_{ab}^{(\vartheta)} n^{a} dS^{b}
  \,.
\end{align}
Here $n^{a}$ is the timelike unit normal to the spatial slice, and
$dS^{b}$ is the proper area element of $S^{2}_{R}$,
i.e.~$dS^{b} = N^{b} \sqrt{\gamma} dA$, where $N^{b}$ is the
spacelike unit normal to $S^{2}_{R}$, $\gamma$ is the determinant
of the induced 2-metric, and $dA$ is the coordinate area element.

Like the metric and scalar field, we similarly expand
$T_{ab}^{(\vartheta)}$ and $\dot E^{(\vartheta)}$ in powers of $\eps$,
\begin{align}
  T_{ab}^{(\vartheta)} =&
\sum_{k=0}^{\infty} \eps^{k} T_{ab}^{(\vartheta,k)}\,, &
  \dot{E}^{(\vartheta)} =&
\sum_{k=0}^{\infty} \eps^{k} \dot{E}^{(\vartheta,k)}\,,
\end{align}
where each $\dot E^{(\vartheta, k)}$ includes the appropriate orders
of both the scalar field and metric.  Since
$\vartheta^{(0)}=0$ and $T_{ab}^{(\vartheta)}$ is quadratic in
$\vartheta$, we have
$T_{ab}^{(\vartheta,0)}=T_{ab}^{(\vartheta,1)}=0$, and similarly
$\dot{E}^{(\vartheta,0)}=\dot{E}^{(\vartheta,1)} = 0$.
The first non-vanishing order is
$T_{ab}^{(\vartheta,2)}$, which is given by
\begin{align}
\label{eq:T-vartheta-2}
T_{ab}^{(\vartheta, 2)} = \nabla_a \vartheta^{(1)} \nabla_b \vartheta^{(1)}
- \frac{1}{2} g_{ab} \nabla_c \vartheta^{(1)} \nabla^c \vartheta^{(1)}
\,.
\end{align}

Using the results of the simulations,
we compute $T_{ab}n^a$, interpolate it onto surfaces of fixed
coordinate radius $R$, compute $T_{ai} n^{a} N^i$ by contracting with the
normal, and perform spectral integration with the induced area element
to obtain $\dot{E}^{(\vartheta,2)}$.  That is, we compute
\begin{align}
\dot{E}^{(\vartheta, 2)} (R) =
\int_{S^2_R} T_{a i}^{(\vartheta, 2)} n^a N^i \sqrt{\gamma} dA
\,.
\label{eq:NumericalScalarFlux}
\end{align}

\figFluxes{}

We also compute the energy flux at order $(\ell/GM)^0$, which for vanishing
$\vartheta^{(0)}$ consists purely of the background gravitational energy flux,
as (c.f.~\cite{Ruiz:2007yx})
\begin{align}
\label{eq:GRFlux}
\dot{E}^{(0)} = \lim_{R \to \infty} \dfrac{R^2}{16 \pi G} \int_{S_R^2} \left| \int_{-\infty}^t \Psi_4 dt' \right|^2 d \Omega\,,
\end{align}
where numerically we set the lower bound of the time integral to the
start of the simulation, assuming there was comparatively little
radiation before the start.

We plot the numerical values of $\dot{E}^{(\vartheta, 2)} (R)$ and
$\dot{E}^{(0)}(R)$ in Fig.~\ref{fig:Fluxes}, keeping
(spin-weighted) spherical harmonics up through $l=8$.
We check for the convergence of the flux quantities with increasing 
extraction radius, and present the results at $R = 300~GM$,
which agree with the results at $R = 200~GM$.

In Fig.~\ref{fig:Fluxes} we also plot a post-Newtonian approximation
to $\dot{E}^{(\vartheta, 2)}$.  This is computed using the far-zone
PN solution for $\vartheta^{(1)}$ from~\cite{Yagi:2011xp}, which only
includes the $l=2$ quadrupole radiation.
We impose circular orbits and aligned spins, convert to our
conventions via Eq.~\eqref{eq:YagiConversion}, and re-insert the
appropriate factors of $G$.  The result for at least one non-zero
spin is
\begin{align}
\label{eq:SpinFluxPN}
 \dot{E}^{(\vartheta, 2)} _\mathrm{PN} =&
-\frac{5}{1536 G}
\left(\frac{\ell}{GM}\right)^4
\left(\tfrac{m_2}{M} \chi_1 - \tfrac{m_1}{M} \chi_2 \right)^2
(GM\omega)^{14/3}
\,,
\end{align}
and for two non-spinning black holes,
\begin{align}
\dot{E}^{(\vartheta, 2)} _\mathrm{PN} =
-\frac{2}{15 G}
\left(\frac{\ell}{GM}\right)^4
\eta^2 \frac{\delta m^2}{M^2}
(GM\omega)^{8}
\,.
\label{eq:NoSpinFluxPN}
\end{align}
In these expressions, $\chi_A$ is the dimensionless spin of black hole
$A$, $\eta = m_1m_2/M^2$ is the symmetric mass ratio, and
$\delta m = m_1 - m_2$ is the mass difference.

Although the gravitational flux at order $(\ell/GM)^0$ is by far the
largest energy flux, the scalar field flux at order $(\ell/GM)^4$
sharply increases before merger.  The spin contributions are dominant,
as the scalar flux for the spin-0 simulation is comparatively small until the
merger, when nonlinearities become very important.  At early
times, our fully numerical results qualitatively agree with the PN
results of~\cite{Yagi:2011xp}, validating our and their calculations.
We expect the $\mathcal{O}(1)$ ratio between PN and full numerics
in Fig.~\ref{fig:Fluxes} stems from the PN
expressions~\eqref{eq:SpinFluxPN}, \eqref{eq:NoSpinFluxPN} only
including $l=2$, whereas our numerics include all modes up through
$l=8$.

\subsection{Regime of validity}
\label{sec:RegimeOfValidity}

Since this method is perturbative, we expect that it breaks
down---becomes invalid---at some point.  There are two types of
breakdown.  First, at every instant of time, there is the question of
whether the series converges.  We expect that the series should only
converge when $\ell \ll GM$, and we assess this in
Sec.~\ref{sec:Epsilon}.  Second, over much longer times, there will be
a secular drift between the perturbative solution and the ``true''
solution, so that the two solutions become out of phase.  We assess
the dephasing below in Sec.~\ref{sec:Dephasing}.

\subsubsection{Instantaneous validity}
\label{sec:Epsilon}

\figInstantPert{}

The perturbative scheme is valid pointwise at every instant in time if
the series for the metric~\eqref{eq:MetricExpansion} and
scalar~\eqref{eq:ScalarFieldExpansion} are convergent.  Roughly, we
can assess this by comparing the magnitudes of successive terms in the
series.  As shown in Sec.~\ref{sec:order_reduction}, up through order
$\eps^{2}$, the metric and scalar are expanded as
\begin{subequations}
  \begin{align}
    g_{ab} ={}& g_{ab}^{(0)} + \eps^{2} h_{ab}^{(2)} + \mathcal{O}(\eps^{3}) \,, \\
    \vartheta ={}& \eps \vartheta^{(1)} +\mathcal{O}(\eps^{3}) \,.
  \end{align}
\end{subequations}
Thus we cannot assess the convergence of $\vartheta$ without going to
$\mathcal{O}(\eps^{3})$, but at $\mathcal{O}(\eps^{2})$ we can
compare the magnitudes of $g_{ab}^{(0)}$ and $h_{ab}^{(2)}$.  A rough
condition for convergence is that
\begin{align}
\label{eq:h2-g0-ineq}
  \left\| h_{ab}^{(2)} \right\| \lesssim
  \left\| g_{ab}^{(0)} \right\|
\,,
\end{align}
where $\left\|\cdot\right\|$ is an $L^{2}$ norm.

The magnitude of $h_{ab}^{(2)}$ depends on the strength of the
coupling parameter $\ell$, as discussed in
Sec.~\ref{sec:summary-scaling}, via
$h_{ab}^{(2)} = (\ell/GM)^{4} \Upsilon_{ab}$, where $\Upsilon_{ab}$ is
independent of $\ell$.  Thus we translate Eq.~\eqref{eq:h2-g0-ineq}
into a condition on the maximum allowed value of $\ell/GM$,
\begin{align}
  \label{eq:ell-over-GM-condition}
  \left|
  \frac{\ell}{GM}
  \right|_{\max} \sim C
  \left(
  \frac{\left\| g_{ab}^{(0)} \right\|}{\left\| \Upsilon_{ab} \right\|}
  \right)^{1/4}_{\min} \,,
\end{align}
where $C$ is some factor of order unity, and on the right-hand side,
the ratio is evaluated pointwise, and then the minimum is taken over
the domain outside of the apparent horizons, at each coordinate time.
At values of $\ell/GM$ larger than this estimate, we expect the
perturbative approach fails to converge somewhere in the spacetime.

In these order $\eps^1$ simulations, we have not simulated
$\Upsilon_{ab}$.  We can, however, make scaling estimates from its
schematic equation of motion, Eq.~\eqref{eq:Upsilon-EOM}.
The source term $\mpl \ell^{2} C_{ab}^{(1)}$ should be of the same
order of magnitude as $T_{ab}^{(2)}$ (which we do compute in the
simulations), so to within an order of magnitude, we estimate
\begin{align}
\square^{(\mathrm{0})} \Upsilon &\sim T_{ab}[\Psi]
\,, \\
\frac{1}{L^2} \| \Upsilon_{ab}^{(2)} \| &\sim \| T_{ab}[\Psi] \|
\,.
\end{align}
Here $L$ is a characteristic curvature length scale, and
$T_{ab}[\Psi]$ is shorthand for the ``stress-energy'' $T_{ab}[\Psi] =
\cd_{a}\Psi \cd_{b} \Psi - \frac{1}{2}g_{ab}(\cd \Psi)^{2}$.
Therefore, we estimate the allowed value for $\ell/GM$ as
\begin{align}
  \label{eq:ell-over-GM-condition-combined}
  \left|
  \frac{\ell}{GM}
  \right|_{\max} \sim C
  L^{-1/2}
  \left(
  \frac{\left\| g_{ab}^{(0)} \right\|}{\left\| T_{ab}[\Psi] \right\|}
  \right)^{1/4}_{\min} \,.
\end{align}

We plot this estimate in
Fig.~\ref{fig:Epsilon} for each of the spin configurations considered
in this study.  During inspiral, the curvature is highest around
the smaller black hole, so we let $L=\min(Gm_{1},Gm_{2})$.  After
merger, we let $L=Gm_{\mathrm{Final}}$ (see Table~\ref{tab:runs} for
values).

We can compare our estimates for the regime of validity
$|\ell/GM|_{\max}$ to those computed in Stein~\cite{Stein:2014xba}.
Stein computed $|\ell/Gm|_{\max}$ of a stationary, isolated black hole
as a function of $\chi$ of the body, using methods that are
independent of ours.  At late times, we find direct agreement, at the
5\% level, by setting $C=(32)^{1/4} \approx 2.38$.  At early times,
after including a factor of $M/m_{2}$ to convert from $|\ell/GM|$ to
$|\ell/Gm_{2}|$, we again find agreement.  At early times, the
low-spin simulation has a very large regime of validity, because the
Pontryagin density is small, and hence Chern-Simons effects are also
small.  However, approaching the time of merger, orbital motion and
nonlinearities source enough energy density in the scalar field to
restrict the regime of validity of $|\ell/GM|$ to order unity.

\subsubsection{Secular validity (dephasing)}
\label{sec:Dephasing}

The true physical system at $\eps>0$ radiates energy more quickly than
the GR-only ($\eps=0$) solution that we are using as the background
for perturbation theory.  As a result, the true solution will inspiral
more quickly, so the orbital phase will have a \emph{secularly}
growing deviation away from the background.  A post-Newtonian scaling
estimate (see below) says that the standard solution will break down
over a secular timescale of order
$T_{\mathrm{sec}} \sim T^{\textrm{GR}}_{\mathrm{RR}} (\ell/GM)^{-2} v^{-2}$,
where $T^{\textrm{GR}}_{\mathrm{RR}}$ is the radiation-reaction
timescale in GR.  This scaling $(\ell/GM)^{-2}$ is characteristic of
singular perturbation theory~\cite{MR538168, Chen:1995ena,
  Galley:2016zee}.

If the length of a detected gravitational waveform is long compared to
the secular breakdown time, then we will need a method to extend the
secular regime of validity of the calculation---for example,
multiple-scale analysis (MSA)~\cite{MR538168} or the dynamical
renormalization group~\cite{Chen:1995ena, Galley:2016zee}.  We save this
issue for future work.  Here, we will estimate the dephasing
time (secular breakdown time).

Let us focus on quasi-circular, adiabatic inspirals.  Similarly to the
scalar field and metric variables in Eqs.~\eqref{eq:MetricExpansion}
and \eqref{eq:ScalarFieldExpansion}, we can expand the accumulated
orbital phase $\phi(t)$ and the orbital frequency
$\omega(t) = \dot{\phi}(t)$ of the binary in powers of $\eps$,
\begin{align}
\phi &= \phi^{(0)} + \eps \phi^{(1)} + \eps^2 \phi^{(2)} + \mathcal{O}(\eps^3)
\,, \\
\omega &= \omega^{(0)} + \eps \omega^{(1)} + \eps^2 \omega^{(2)} + \mathcal{O}(\eps^3)
\,,
\end{align}
where $\phi^{(0)}$ corresponds to the phase of the binary in pure GR,
and $\phi^{(1)}$ contains the dCS corrections at order $\eps^1$ and so
on.  Since the metric deformation at $\mathcal{O}(\eps^{1})$ vanishes,
the phase correction at $\mathcal{O}(\eps^{1})$ also vanishes,
$\phi^{(1)}=0=\omega^{(1)}$.  The first non-vanishing orbital phase
correction is
\begin{align}
\Delta \phi \equiv \phi^{(2)}
\,.
\end{align}
We can use $\Delta \phi$ to assess the secular regime of
validity, and in Sec.~\ref{sec:Detectability} we will also use it to
assess the detectability of dynamical Chern-Simons.

We do not have $\Delta \phi$ directly from the simulation, as we do
not evolve the $\eps^2$ system.  However, we can estimate it from
previously-known analytical results combined with numerical quantities
available during the simulation.

Consider the local-in-time expansion of the orbital phase correction
$\Delta\phi$ around any `alignment time' $t_{0}$,
\begin{align}
  \label{eq:Delta-phi-expansion}
  \Delta\phi(t) ={}&
  \Delta\phi(t_{0})
  + (t-t_{0}) \frac{d \Delta\phi}{dt} \Big|_{t=t_{0}}
  \\
  &{}+ \frac{1}{2} (t-t_{0})^{2} \frac{d^{2} \Delta\phi}{dt^{2}} \Big|_{t=t_{0}}
  + \mathcal{O}(t-t_{0})^{3}
  \,, \nonumber \\
  \label{eq:Delta-phi-expansion-omegas}
  \Delta\phi(t) ={}&
  \Delta\phi(t_{0})
  + (t-t_{0}) \omega^{(2)}(t_{0})
  \\
  &{}+ \frac{1}{2} (t-t_{0})^{2} \frac{d \omega^{(2)}}{dt} \Big|_{t=t_{0}}
  + \mathcal{O}(t-t_{0})^{3}
  \,. \nonumber
\end{align}
If our simulation had started at reference time $t_{0}$, then we would
have $\Delta\phi(t_{0}) = 0$.
The linear piece $(t-t_{0}) \omega^{(2)}(t_{0})$ corresponds to a
perturbative, instantaneous frequency shift, which is completely
degenerate with a renormalization of the physical mass $M(\eps)$
in terms of the `bare' mass $M(\eps = 0)$.  Therefore, the constant
and linear pieces of this expansion are not observable.

However, the curvature
$\frac{1}{2} (t-t_{0})^{2} d\omega^{(2)}/dt|_{t=t_{0}}$ cannot be
redefined or scaled away.  Therefore, within a sufficiently short
window of time around the alignment time $t_{0}$, the deformation to
the orbital phase is given by
\begin{align}
\label{eq:Delta-phi-approx}
  \Delta\phi =
  \frac{1}{2} (t-t_{0})^{2} \frac{d \omega^{(2)}}{dt} \Big|_{t=t_{0}}
  + \mathcal{O}((t-t_{0})^{3})
  \,.
\end{align}
We use this to define the perturbative secular time
$T_{\mathrm{sec}}(t_{0})$ at any instant $t_{0}$ via
\begin{align}
  1 \approx \Delta\phi &=
  \frac{1}{2} T_{\mathrm{sec}}^{2} \frac{d \omega^{(2)}}{dt} \Big|_{t=t_{0}}
  \,, \\
  \label{eq:T-sec-def}
  T_{\mathrm{sec}} &\equiv
  \left( \frac{1}{2} \frac{d\omega^{(2)}}{dt} \Big|_{t=t_{0}} \right)^{-1/2} \,,
\end{align}
roughly the time to dephase by order one radian.

Thus we need to estimate $d\omega^{(2)}/dt$ from our simulation.
Under the assumption of quasi-circular, adiabatic orbits, there is a
one-to-one correspondence between the orbital frequency $\omega$ and
orbital energy $E$.  In other words, there exists a function of one
variable, $E(\omega)$ or $\omega(E)$.  Therefore from the chain rule
we can find the time derivative
\begin{align}
  \label{eq:chain-rule}
  \frac{d\omega}{dt} = \frac{d\omega}{dE} \frac{dE}{dt}
  = \frac{dE/dt}{dE/d\omega} \,.
\end{align}
This depends on the conservative sector through the
frequency-dependence of orbital energy, $dE/d\omega$, and on the
dissipative sector through the radiated power, $dE/dt$.  Just as with
the frequency, we expand the orbital energy in powers of $\eps$,
\begin{align}
E = E^{(0)} + \eps E^{(1)} +
\eps^2 E^{(2)} + \mathcal{O}(\eps^3)
\,.
\end{align}
We can then use this to expand Eq.~\eqref{eq:chain-rule} in powers of
$\eps$.  The $\mathcal{O}(\eps^{2})$ piece is given by
\begin{align}
  \label{eq:domega2/dt}
  \frac{d\omega^{(2)}}{dt} = \frac{d\omega^{(0)}}{dt}
  \left[
  \frac{dE^{(2)}/dt}{dE^{(0)}/dt} - \frac{dE^{(2)}/d\omega}{dE^{(0)}/d\omega}
  \right] \,.
\end{align}
The prefactor $d\omega^{(0)}/dt$ is simply the background (GR)
evolution of the orbital frequency.  The first term in square brackets
in Eq.~\eqref{eq:domega2/dt} comes from the dissipative sector of the
dynamics, since it depends on the radiated power $dE^{(2)}/dt$.  The
second term, meanwhile, comes from the conservative sector, as it
depends on the correction to the orbital energy $E^{(2)}(\omega)$.
Both of the factors in square brackets scale as
$(\ell/GM)^{4} v^{4}$~\cite{Yagi:2011xp, Yagi:2013mbt} for BBHs with
spin.  Plugging this scaling into Eq.~\eqref{eq:T-sec-def} recovers
$T_{\mathrm{sec}} \sim T^{\textrm{GR}}_{\mathrm{RR}} (\ell/GM)^{-2} v^{-2}$.

We find it useful to rewrite $dE^{(0)}/d\omega$ in the second term
using the chain rule~\eqref{eq:chain-rule} to give
\begin{align}
  \label{eq:domega2/dt-final}
  \frac{d\omega^{(2)}}{dt} =
  \frac{d\omega^{(0)}/dt}{dE^{(0)}/dt}
  \left[
  \frac{dE^{(2)}}{dt} -
  \frac{d\omega^{(0)}}{dt} \frac{dE^{(2)}}{d\omega}
  \right] \,.
\end{align}
Now we can discuss how to evaluate these factors from our numerical
simulation and previously-known analytical results.  The background
energy flux $dE^{(0)}/dt$ comes from the numerical simulation via
Eq.~\eqref{eq:GRFlux}.  We also have the background frequency
evolution $d\omega^{(0)}/dt$ from the numerical simulation, via a
time derivative of the coordinate orbital frequency.

The two $\mathcal{O}(\eps^{2})$ quantities require approximations.
In the dissipative sector, there are two contributions to
$dE^{(2)}/dt$: the first from scalar radiation, and the second from
gravitational radiation.  We expect these to be the same order of
magnitude.  Since we do not have access to the gravitational
radiation, we approximate that to within an order of magnitude,
\begin{align}
  \label{eq:dot-E-2-approx}
  \dot{E}^{(2)} \approx \dot{E}^{(\vartheta,2)}
  \,,
\end{align}
where $\dot{E}^{(\vartheta,2)}$ was given in
Eq.~\eqref{eq:NumericalScalarFlux}.
This is further justified during the inspiral, where the
$\mathcal{O}(\eps^{2})$ dissipative correction due to gravitational
waves is higher-PN than the scalar radiation~\cite{Yagi:2011xp}.

In the conservative sector, we can approximate $E^{(2)}(\omega)$ from
a post-Newtonian calculation~\cite{Yagi:2013mbt, Stein:2013wza}.  The
(PN-approximate) correction to the orbital energy $E^{(2)}$ also has
two pieces: the scalar binding energy and the metric-deformation
binding energy.  Again we are going to make an approximation and
ignore the metric deformation piece, approximating
\begin{align}
  \label{eq:E-2-of-omega-approx}
  E^{(2)}(\omega) \approx E^{(\vartheta,2)}_{\mathrm{DD}}
  \,,
\end{align}
where $E^{(\vartheta)}_{\mathrm{DD}}$ is the scalar dipole-dipole
interaction.  After accounting for a missing minus sign
in~\cite{Yagi:2013mbt, Stein:2013wza}, this is given by
\begin{align}
E^{(\vartheta,2)}_\mathrm{DD} &=
  4 \pi \frac{3 \mu_{1}^{i} \mu_{2}^{j} n^{12}_{\langle ij\rangle }}{r_{12}^{3}}
  \\
  &=
  \frac{4 \pi}{r_{12}^3}
  \left[
  3 (\mu_1 \cdot n_{12})(\mu_2 \cdot n_{12}) -
  (\mu_1 \cdot \mu_2)\right]
\,,
\end{align}
where again $\mu_A^{i}$ is the scalar dipole moment given in
Eq.~\eqref{eq:scalar-dipole-moment}.
In our case the spins are in the $\hat z$ direction, so the
$(\mu_{A} \cdot n_{12})$ term vanishes.
To leading PN order, we use the Kepler relation $\omega^{2} =
GM/r_{12}^{3}$ and obtain
\begin{align}
E^{(\vartheta,2)}_\mathrm{DD} &= 4 \pi \omega^{2} (GM)^{-1} \mu_1 \mu_2 \\
\frac{d E^{(\vartheta,2)}_\mathrm{DD}}{d \omega} &= 8 \pi \omega (GM)^{-1} \mu_1 \mu_2
\,,
\end{align}
where $\mu_A$ now refers to the $\hat z$ component.
For $\omega$ we again use the coordinate orbital frequency from the
simulation.

To summarize this calculation: we are approximating the secular
breakdown time $T_{\mathrm{sec}}$ [Eq.~\eqref{eq:T-sec-def}] by
assuming a quasi-circular, adiabatic inspiral, and thus we compute
$d\omega^{(2)}/dt$, Eq.~\eqref{eq:domega2/dt-final}.  We approximate
the dissipation $\dot{E}^{(2)}$ from only the scalar flux,
Eq.~\eqref{eq:dot-E-2-approx}.  We approximate the conservative
correction $E^{(2)}(\omega)$ from the post-Newtonian scalar
dipole-dipole interaction,
Eq.~\eqref{eq:E-2-of-omega-approx}.

\figDephasingWindow{}

In Fig.~\ref{fig:DephasingTime} we plot
$(\ell/GM)^{2}T_{\mathrm{sec}}(t_{0})$, the time to secularly dephase
by about $\sim 1$ radian, around various alignment times $t_{0}$.  We
have checked that at early times, this numerical estimate agrees with
an analytic PN estimate.  As expected, $T_{\mathrm{sec}}$ is
parametrically longer than the GR radiation-reaction time.  The time
window for secular validity shrinks approaching merger, but does not
vanish.

The value of $T_{\mathrm{sec}}$, and hence secular regime of validity,
is smallest near merger.  For the spin 0.3 simulation, just before merger, we
find the time to dephase by about 1 radian from the GR background is
$T_{\mathrm{sec}}\sim 15~GM (\ell/GM)^{-2}$.  If Advanced LIGO detects
a gravitational waveform of length, say, $200~GM$, then a perturbative
calculation without MSA/renormalization would be valid for
$(\ell/GM) \lesssim 1/4 $.  For longer waveforms or larger values of
$(\ell/GM)$, MSA or renormalization would be required.  However,
larger values of $(\ell/GM)$ will be very close to the limit on the
instantaneous regime of validity, Fig.~\ref{fig:Epsilon}.

\subsection{Detectability and bounds estimates}
\label{sec:Detectability}

We now turn to the issue of how well Advanced LIGO/Virgo would be able
to detect or bound the effects of dynamical Chern-Simons gravity
from observations of a binary black hole merger.  As we do not yet have metric
waveforms [that arise at $\mathcal{O}(\eps^{2})$], we make
order-of-magnitude projections of detectability and bounds from the
dephasing estimates in the previous section.

Suppose that LIGO detects a gravitational waveform similar to one of
those we have simulated, with approximately 5 cycles of inspiral in
band before merger---similar to GW150914~\cite{Abbott:2016blz}, with a
total mass approximately $M\approx 60 M_{\odot}$.  Such a detection
would come with errors due to noise and calibration uncertainty; let
us define the overall waveform phase uncertainty $\sigma_{\phi}$.
Let us further assume that the dCS corrections to the full waveform are not
degenerate with redefining `bare' binary parameters.
Upon detection there are two distinct possibilities: (i) the detected
waveform is consistent with GR predictions; or (ii) the detection is
inconsistent with any point in the GR parameter space.

In the case of consistency, we would be able to place bounds on
the size of $\ell$.  Crudely, we would be able to say
\begin{align}
  \Delta\phi_{\textrm{gw}} = 2\Delta\phi \lesssim \sigma_{\phi}
  \,,
\end{align}
where the factor of two comes from the gravitational wave being at
twice the orbital frequency.  If we have consistency with GR, then the
quadratic approximation for $\Delta\phi$ in
Eq.~\eqref{eq:Delta-phi-approx} holds.

\figDephasing{}

We plot the quadratic approximation to the orbital phase difference
(relative to GR) in Fig.~\ref{fig:Dephasing}.  By taking the maximum
value of $\Delta\phi$ over the length of the waveform, and taking into
account the scaling with $(\ell/GM)^{4}$, we can derive a projected
bound on $\ell$.  For example, from the spin 0.3 simulation and
$M\approx 60 M_{\odot}$, we would find
\begin{align}
  \left( \frac{\ell}{GM} \right) &
  \lesssim 0.13 \left( \frac{\sigma_{\phi}}{0.1} \right)^{1/4}
&\textrm{or}
&&
  \ell &
  \lesssim 11~\textrm{km} \left( \frac{\sigma_{\phi}}{0.1} \right)^{1/4}
\,,
\end{align}
and from the spin 0.1 simulation,
\begin{align}
  \left( \frac{\ell}{GM} \right) &
  \lesssim 0.2 \left( \frac{\sigma_{\phi}}{0.1} \right)^{1/4}
&\textrm{or}
&&
  \ell &
  \lesssim 18~\textrm{km} \left( \frac{\sigma_{\phi}}{0.1} \right)^{1/4}
\,.
\end{align}
The spin 0.0 simulation would only give
$(\ell/GM) \lesssim 1.4 (\sigma_{\phi}/0.1)^{1/4}$.  Such a bound
would be past the instantaneous regime of validity limit during merger
for this simulation (see Fig.~\ref{fig:Epsilon}).  It is not internally
self-consistent to use this perturbative result to claim a constraint
on the regime past perturbative validity, so conservatively, no
statement can be made.  The higher spin simulations do not suffer from this
problem.

These bounds forecasts can immediately be turned around into
detectability forecasts.  We can forecast that dynamical Chern-Simons
corrections would be detectable in a $M\approx 60 M_{\odot}$ binary
with parameters consistent with our spin 0.3 simulation if
$\ell \gtrsim 11$~km, and similarly for the spin 0.1 simulation if
$\ell \gtrsim 18$~km.

We can draw three simple lessons on detectability and bounds from
these results.  First, better phase sensitivity (smaller
$\sigma_{\phi}$) is an obvious way to improve the odds of
detectability, or place stronger bounds.  This comes from improved
detector sensitivity, but also from higher signal-to-noise ratio (SNR)
events.  Second, at fixed phase sensitivity, lower-mass events would
be better than higher mass events, to a point.  Lower mass events
obviously have smaller $GM$, but they also spend more time in band,
and thus have more time for dephasing.  There is a tradeoff, though,
because lower mass events are quieter, and also because most of the
dephasing comes right before merger---so the mass must be high enough
for merger to be in band.  Finally, we can easily see that higher spin
systems would lead to stronger constraints or a better chance of
detecting dCS effects.

Let us compare our projected bounds to those appearing previously in
the literature.
Ali-Haïmoud and Chen~\cite{AliHaimoud:2011fw} used solar system data
from Gravity Probe B and the LAGEOS satellites to constrain the
characteristic length scale to $\ell \lesssim 10^8~\mathrm{km}$.
Yagi, Yunes and Tanaka~\cite{Yagi:2012ya} found a similar bound from
table-top experiments. This is comparable to the curvature radius in
the solar system.

Yunes and Pretorius~\cite{Yunes:2009hc} applied a precession
calculation from the extreme mass-ratio limit to PSR~J0737--3039 to
estimate a constraint of $\ell \lesssim 10^4~\mathrm{km}$.  However,
this calculation missed some effects (such as the scalar binding
energy), and the mass ratio of PSR~J0737--3039 is very close to 1.
Moreover, the curvature radius at the surface of one of the NSs in
this system should be order $\sim 10$~km, which means there is room
between $10-10^{4}$~km where $\ell$ could be large compared to the
curvature length, and thus the calculation would not be internally
self-consistent.
Yagi, Stein, Yunes, and Tanaka~\cite{Yagi:2013mbt} performed a more
careful analysis, using post-Newtonian theory for binary NS systems.
They concluded that even PSR~J0737--3039, with its high orbital
velocity and exquisite timing, would not be able to yield a constraint
on dCS for the foreseeable future, and that gravitational wave
measurements would be the best hope.

Yagi, Yunes, and Tanaka~\cite{Yagi:2012vf} used post-Newtonian
calculations to project the level of constraints that might arise from
second and third generation GW detectors.  If next-generation
detectors such as Einstein Telescope~\cite{Punturo:2010zza} were to
observe binary black hole inspirals consistent with GR, then YYT
project a bound of $\ell \lesssim \mathcal{O}(10 -100)~\mathrm{km}$.
Second-generation ground-based detectors could place a similar
constraint.  The only caveat here is that YYT use post-Newtonian
estimates, stopping at the ISCO frequency, for systems that would be
seen not only in the inspiral, but also in the merger and ringdown,
where PN is invalid.  The additional SNR contributed by merger and
ringdown will likely improve constraints.

Stein and Yagi~\cite{Stein:2013wza} projected a number of constraints
on $\ell$ based on both pericenter precession in pulsar binaries and
gravitational wave measurements.  For a LIGO detection of a
$(10+11) M_{\odot}$ BBH inspiral, consistent with GR, at an SNR of 30,
they projected a bound on the order of $\ell \lesssim 10~\mathrm{km}$.
Note that this is the same order of magnitude as the projected bound
we estimate here.

Finally, Stein~\cite{Stein:2014xba} projected a bound based on the
observations of the black hole candidate GRO~J1655--40.
Assuming observations were consistent with GR, Stein projected a
constraint of $\ell \lesssim 22~\mathrm{km}$.  However, such a
constraint would require (for example) accretion disk modeling in the
presence of the dCS correction, which has not been simulated.

\section{Discussion and future work}
\label{sec:discussion}

In this study, we have performed the first fully nonlinear inspiral,
merger, and ringdown numerical simulations of a binary black
hole system in dynamical Chern-Simons gravity.  These are the first
BBH simulations in a theory besides general relativity and
standard scalar-tensor gravity.  BBH in scalar-tensor is identical to
that in GR, unless one imposes an external scalar field
gradient~\cite{Healy:2011ef, Berti:2013gfa}.  Therefore these are also
the first numerical simulations in a theory where the BBH dynamics
differ from GR under ordinary initial and boundary conditions.

The ``full'' equations of motion for dCS, and many other corrections
to GR, probably lack a well-posed initial value formulation.  This is
not an obstacle if the corrections are treated as being a low-energy
effective field theory.  In Sec.~\ref{sec:formalism} we formulated a
perturbation scheme which guarantees a well-posed initial value
problem.  We stress that this scheme is applicable not just to
dCS, but also any deformation of general relativity which has a
continuous limit to GR.

We performed fully nonlinear numerical simulations through order
$\mathcal{O}(\eps^{1})$ in the perturbation scheme.  We simulated
binaries with mass ratio $q=3$ and aligned spins with equal
dimensionless spin parameters $\chi_{1}=\chi_{2}$, taking on three
values, $\chi=0.0, 0.1, 0.3$.  The background ($\eps^{0}$) metric
radiation and perturbative ($\eps^{1}$) scalar radiation waveforms are
presented in Sec.~\ref{sec:Waveforms}.  We found good agreement with
PN waveform predictions~\cite{Yagi:2011xp, Yagi:2013mbt} during the
early inspiral.

We have also discovered new phenomenology in dCS.  In agreement with
PN predictions, dCS does not suffer from dipole radiation during the
early inspiral.  However, during merger, there is a burst of dipole
radiation.  This phenomenon can only be studied with full numerical
simulations.

We extracted energy fluxes in Sec.~\ref{sec:EnergyFluxes}, finding
good agreement with PN at early times.  We found that the scalar
field's $\mathcal{O}(\eps^{2})$ energy flux during the inspiral was
approximately $10^{-6}(\ell/GM)^{4}$ times smaller than the
corresponding $\mathcal{O}(\eps^0)$ (GR) energy flux for the highest
spin simulation, rising to a $10^{-3}(\ell/GM)^{4}$ fraction of GR
during merger.  This energy flux enters into our detectability
estimate.

Since we use a perturbative scheme, it is important to understand
where perturbation theory breaks down.  In Sec.~\ref{sec:Epsilon} we
estimated the maximum values of $\ell/GM$ for the perturbation theory
to be convergent at each time during the simulation.  During the
inspiral and ringdown, the regime of validity agrees with estimates
from~\cite{Stein:2014xba}.  The tightest bound on the
instantaneous regime of validity comes during merger, and is
comparable for spinning and non-spinning black hole mergers, close to
$\ell/GM \lesssim 1$.

The additional radiation in the scalar field $\vartheta^{(1)}$ leads
to a secular drift in orbital phase between the ``true'' orbital
dynamics and the GR background from which we perturb.  Therefore even
if perturbation theory is instantaneously under control, the
perturbative solution will dephase after a sufficiently long time.  We
numerically estimated this dephasing time in Sec.~\ref{sec:Dephasing},
and it agrees with post-Newtonian scaling at early times.  At times
approaching merger, the dephasing time becomes shorter, but remains
nonzero.

This dephasing calculation served as the basis for estimating
detectability and predicting bounds that LIGO would be able to
place on $\ell$, in Sec.~\ref{sec:Detectability}.  For $q=3$,
$M\approx 60M_{\odot}$, and aligned dimensionless spins of
$\chi_{1}=\chi_{2}=0.3$, we estimated that a GR-consistent detection
would yield a bound of
\begin{align}
  \ell \lesssim 11~\textrm{km} \left( \frac{\sigma_{\phi}}{0.1} \right)^{1/4}
  \,,
\end{align}
where $\sigma_{\phi}$ is LIGO's statistical phase uncertainty on the
detected waveform, which depends on the SNR of
the detection.  Conversely, an $\ell$ above this value would be
detectable by LIGO.  Lower spins lead to poorer detectability and/or
bounds.  Better bounds come from three places: (i) improved phase
sensitivity (higher SNR), (ii) lower mass events (while keeping merger
in band), and (iii) higher spin systems.

\subsection{Future work}
\label{sec:future-work}

The natural next step in this program is to continue to the order
$\eps^{2}$ system, as outlined in Sec.~\ref{sec:ordereps2}.  This is
the lowest order where gravitational radiation is modified, and would
involve solving for $h_{ab}^{(2)}$, which is sourced by $g_{ab}^{(0)}$
and $\vartheta^{(1)}$.

With the solution for the deformation to the metric $h_{ab}^{(2)}$, we
will be able to directly compare dCS predictions against LIGO data.
We will also have a more complete assessment of the convergence of the
perturbation scheme.

Comparing dCS predictions against LIGO data will yield the first
direct bounds on the theory from the strong-field, dynamical regime of
gravity.  To do so will involve extending GR parameter
estimation~\cite{PhysRevD.91.042003} with one additional parameter,
$\ell$, which will be simultaneously inferred or constrained from the
data.

A complete analysis would involve thorough exploration of the
7-dimensional parameter space of quasicircular BBHs (mass ratio and
two spin vectors; the $\ell$ dependence is analytic in the
perturbative approach).  For example, in this work, we have focused on
aligned-spin binaries in order to simplify comparisons with
analytic predictions.  The scalar energy flux in the case of
misaligned binaries may be an order of magnitude larger than in the
spin aligned case (see~\cite{Yagi:2011xp} and the erratum).  Building
a surrogate waveform model~\cite{Blackman:2015pia, Blackman:2017dfb}
would simultaneously allow for an efficient exploration of parameter
space and efficient parameter estimation/constraints with LIGO data.

The standard perturbation theory approach we used here will be
sufficient if we find that the dephasing time is long compared to LIGO
signals.  However, if we need to extend the secular regime of
validity, some form of multiple-scale analysis~\cite{MR538168} or
dynamical renormalization group~\cite{Chen:1995ena, Galley:2016zee}
approach will be required.

Finally, let us emphasize that our approach is not limited to
dynamical Chern-Simons gravity: dCS is a proof of principle.  Any
theory with a continuous limit to GR can be treated with the same
scheme, and reusing a large fraction of the code.  In particular, we
will consider EdGB and a class of theories proposed
in~\cite{Endlich:2017tqa}.  Switching from dCS to another theory will
only involve changing source terms that appear on the right hand sides
of the differential equations we are solving numerically.

\acknowledgments

We would like to thank
Yanbei Chen,
Chad Galley,
Luis Lehner,
Robert McNees,
Frans Pretorius,
Thomas Sotiriou,
Saul Teukolsky,
Helvi Witek,
Kent Yagi,
and
Nico Yunes
for many valuable conversations.
This work was supported in part by the Sherman Fairchild Foundation,
the Brinson Foundation,
and NSF grants PHY--1404569 and AST--1333520 at Caltech.
MO gratefully acknowledges the support of the Dominic Orr Graduate
Fellowship at Caltech.
Computations were performed on the Zwicky cluster at Caltech, which is
supported by the Sherman Fairchild Foundation and by NSF award
PHY--0960291.
Some calculations used the computer algebra system
\codename{Mathematica}, in combination with the
\codename{xAct/xTensor} suite~\cite{xActWebPage, MartinGarcia2008597,
  Brizuela:2008ra}.
The figures in this paper were produced with
\codename{matplotlib}~\cite{Hunter:2007}.

\appendix{}
\section{Scalar field evolution formulation}
\label{appendix:ScalarField}

In this appendix, we discuss the numerical evolution scheme for a
(massless) Klein-Gordon field, denoted by the code variable $\Psi$, in
greater detail.  This is an update of the system described
in~\cite{Holst:2004wt}, which did not include the
``$\gamma_{1}\gamma_{2}$'' constraint-damping term (see below).  The
basic equation we are simulating is
\begin{align}
  \label{eq:KG-equation}
  \square \Psi = S \,,
\end{align}
for some prescribed source term $S$ (in this work, the source term is
the Pontryagin density $\pont$).

We first review the 3+1 ADM formalism for the foliation of a spacetime
into spatial slices, as used in numerical
relativity~\cite{baumgarteShapiroBook}.  We decompose the metric as
\begin{align}
g_{ab} = \gamma_{ab} - n_{a} n_{b} \,,
\end{align} 
where $g_{ab}$ is the spacetime metric, $n_a$ is a timelike unit
one-form normal to the spatial slice with $n_{a}n^{a} = -1$, and
$\gamma_{ab}$ is the induced spatial metric and projector, with
$n^{a}\gamma_{ab} = 0$.  In ADM variables, the timelike unit normal
can be written in terms of a lapse, $\alpha$, and shift $\beta^i$, as
$n^a = (\alpha^{-1}, -\alpha^{-1} \beta^i)$.

We work with the Spectral Einstein Code (\codename{SpEC}), which uses
the generalized harmonic formulation of general relativity, and
evolves a symmetric hyperbolic first-order system of metric variables
$g_{ab}$, $\Phi_{iab} = \partial_i g_{ab}$ and
$\Pi_{ab} = -n^c \partial_c g_{ab}$~\cite{Lindblom2006}.

We similarly define a set of first-order variables for the scalar
field $\Psi$ as
\begin{align}
\Phi_i &= \partial_i \Psi \,, \\
\Pi &= -n^a \partial_a \Psi = -\alpha^{-1}(\partial_t \Psi - \beta^i \partial_i \Psi) \,.
\end{align}
From these definitions and the equality of mixed partial derivatives,
we can create a system of constraints which vanish in the continuum
limit, and which an accurate
evolution of the system will satisfy to within some tolerance:
\begin{align}
C^{(1)}_i &= \partial_i \Psi - \Phi_i \label{eq:KGC1} \,, \\
C^{(2)}_{i} &= [ijk] \partial_j \Phi_k \,.
\label{eq:KGC2}
\end{align}
In Eq.~\eqref{eq:KGC2} the indices $j,k$ are summed and $[ijk]$
is the completely antisymmetric Levi-Civita \emph{symbol}, with $[123]=+1$.

The evolution equation~\eqref{eq:KG-equation} thus becomes a set of
first-order time evolution equations for $\{\Psi$, $\Phi_i, \Pi\}$.
However, in order to prevent numerical errors in the constraints from
making the evolution unstable, we follow what is done in the metric
system and add specific multiples of the constraints to the evolution
equations.  These combinations of constraints are chosen so as to
ensure that the system is symmetric hyperbolic and that the
constraints are damped out, ensuring a well-posed evolution scheme.
The evolution equation for $\Psi$ is thus
\begin{align}
\partial_t \Psi =& -\alpha \Pi + \beta^m [ \partial_m \Psi
                   + \gamma_1 (\partial_m \Psi - \Phi_m) ] \,,
\end{align}
where the first terms come from the definitions of $\Phi_i$ and $\Pi$,
and the last term is a constraint damping term with coefficient
$\gamma_1$. The evolution equation for $\Phi_i$ is
\begin{align}
\begin{split}
\partial_t \Phi_k =& - \alpha[\partial_k \Pi + \gamma_2(\Phi_k - \partial_k \Psi)]  \\
&{}- \Pi \partial_k \alpha + \beta^m \partial_m \Phi_k +
\Phi_m \partial_k \beta^m \,,
\end{split}
\end{align}
where the term with $\gamma_2$ is a constraint damping term, and
all other terms come from definitions of the first-order variables and
equality of mixed partial derivatives. Finally, the evolution equation
for $\Pi$ is
\begin{align}
\begin{split}
\partial_t \Pi ={}& \alpha \Pi K + \beta^m \partial_m \Pi + \alpha \Phi_m \Gamma^m \\
&{}+ \gamma_1 \gamma_2 \beta^m (\partial_m \Psi - \Phi_m) \\
&{}- \alpha g^{mn} \partial_n \Phi_m - g^{mn}  \Phi_n \partial_m \alpha \\
&{}+ \alpha S \,,
\end{split}
\end{align}
where $K$ is the trace of the extrinsic curvature,
$\Gamma^{m}\equiv g^{ab}\Gamma^{m}{}_{ab}$ is a specific contraction
of the Christoffel connection coefficients, $S$ is the source term,
and the $\gamma_1 \gamma_2$ term is the appropriate constraint-damping
term to keep the system symmetric hyperbolic.

This ``$\gamma_{1}\gamma_{2}$'' term was not included in the previous
description~\cite{Holst:2004wt}, but it is required if both
$\gamma_{1}$ and $\gamma_{2}$ are non-zero.  The parameters
$\gamma_{1}$ and $\gamma_{2}$ play the same role in the
damping and characteristic analysis of this Klein-Gordon system as
they do in the generalized harmonic system~\cite{Lindblom2006}.
Specifically, in order for the constraint $C_{i}^{(1)}$ to be damped,
we must have $\gamma_{2} > 0$ (satisfying the constraint $C_{i}^{(1)}$
implies satisfaction of the constraint $C_{i}^{(2)}$).  The choice
$\gamma_{1} = -1$ makes the system linearly degenerate.  In practice
we set the values of $\gamma_{1}$ and $\gamma_{2}$ to match those of
the generalized harmonic evolution of the metric variables, so that
the characteristic speeds of the metric and scalar field systems
agree.

The scalar field variables, like the metric variables, are
represented spectrally. In order to reduce the amount of numerical
noise in the system, we apply the same filters we use for the
metric variables to the scalar field system, namely filtering the
top $4$ tensor spherical harmonics and using an exponential Chebyshev
filter for the radial piece.

\figConstraints{}

In order to assess the accuracy of the simulations, we evaluate the
constraints that the generalized harmonic evolution system must
satisfy~\cite{Lindblom2006}, as well as the constraints for
the first-order scalar field system given by Eqs.~\eqref{eq:KGC1}
and~\eqref{eq:KGC2}. We combine these constraints, contracting with a
Euclidean metric to give a constraint
energy as
\begin{align}
C^2 = C^{(1)}_i  C^{(1)}_i +  C^{(2)}_j C^{(2)}_j \,.
\label{eq:KGCE}
\end{align}
Since the code is spectral, we check for exponential convergence of
these constraint energies as we increase the number of angular and
 radial basis functions per subdomain (and hence the resolution). We plot the
convergence of the $L^\infty$ norm of the constraint energies for
the highest spin simulation of this study, which has the greatest level of constraint
violation, in Fig.~\ref{fig:0p3_0p3_Constraint}.  We find that the
error decreases exponentially with resolution. The lower spin simulations
have similar qualitative behavior.

\section{Pontryagin density in 3+1 split}
\label{appendix:EBDecomposition}

Since numerical relativity computations are formulated in a 3+1 split,
we must compute the scalar field's source term---the Pontryagin
density---in terms of 3 dimensional quantities.  First, it is
straightforward to verify
\begin{align}
  \label{eq:CP-identity}
  \pont \equiv {}^{*}\!R^{abcd}R_{abcd} = {}^{*}\!C^{abcd}C_{abcd} \,,
\end{align}
where $C_{abcd}$ is the Weyl tensor, and its left dual is
${}^{*}C^{abcd}\equiv \frac{1}{2} \epsilon^{abef}C_{ef}{}^{cd}$.  Thus
we do not need to consider all of Riemann, but only its trace-free
part, Weyl.  The Pontryagin density is completely insensitive to the
Ricci part of curvature.

In a 4-dimensional numerical relativity simulation, it is especially
convenient to decompose Weyl into its electric and magnetic parts,
defined as
\begin{align}
  \label{eq:E-def}
  E_{ab} &\equiv + {}^{\phantom{*}}\! C_{acbd} n^{c} n^{d}\,, \\
  \label{eq:B-def}
  B_{ab} &\equiv - {}^{*}\!C_{acbd} n^{c} n^{d} \,.
\end{align}
The minus sign in~\eqref{eq:B-def} follows the conventions
of~\cite{OwenEtAl:2011, Nichols:2011pu} and the implementation in
\codename{SpEC}~\cite{SpECwebsite}, though much of the literature has a plus
sign.  From the symmetries of Weyl, the two tensors $E_{ab}$ and
$B_{ab}$ are both symmetric ($E_{ab} = E_{(ab)}$ and
$B_{ab} = B_{(ab)}$), purely spatial
($E_{ab}n^{a} = 0 = B_{ab}n^{a}$), and trace-free
($E^{a}{}_{a} = 0 = B^{a}{}_{a}$).  We may also write an inversion
formula for Weyl in terms of $E_{ab}$ and $B_{ab}$ (thanks to Alfonso
García-Parrado for bringing this inversion formula to our attention),
\begin{align}
  \label{eq:Weyl-from-E-B}
  C_{abcd} = {\raisebox{-0.25em}{\tiny\young(ac,bd)}}
  \left[
  4 E_{ac} (\gamma_{bd}+n_{b}n_{d}) - \epsilon_{ab}{}^{e} n_{d} B_{ce}
  \right]\,,
\end{align}
where the operator ${\tiny\young(ac,bd)}$ is a projector that imposes
the symmetries of the Riemann tensor ($R_{abcd}=R_{[ab][cd]} =
R_{cdab}$).  Here we have the induced 3-dimensional volume element,
\begin{align}
  \label{eq:3-vol}
  \epsilon_{abc} &\equiv n^{d} \epsilon_{dabc} \,, &
  \epsilon_{abcd} &= -4 n_{[a} \epsilon_{bcd]} \,.
\end{align}
For coordinate component calculations, we use the conventions where
$\epsilon_{abcd} = +\sqrt{-g}[abcd]$ where $[abcd]$ is the alternating
\emph{symbol}, with $[0123] = +1$ (see e.g.~\cite{MTW}).  We also have
$\epsilon^{abcd} = -[abcd]/\sqrt{-g}$, and similar conventions for the
3-dimensional volume element: $\epsilon_{abc} = \sqrt{\gamma}[abc]$
and $\epsilon^{abc} = [abc]/\sqrt{\gamma}$ (this makes use of the
identity $\sqrt{-g} = \alpha \sqrt{\gamma}$).

With this above decomposition, it is easy to verify that the
Pontryagin density can be expressed simply in terms of the
electric and magnetic parts of Weyl,
\begin{align}
  \label{eq:pont-from-EB}
  \pont = -16 E_{ab} B^{ab} \,.
\end{align}

Thus all that remains is to compute $E_{ab}$ and $B_{ab}$ from other
quantities. Finding these expressions for $E$ and $B$ comes from the
standard Gauss-Codazzi-Mainardi (GCM) equations
(see~\cite{baumgarteShapiroBook} for a didactic explanation).  After
using the GCM equations, for the electric Weyl tensor we find
\begin{align}
\label{eq:E-from-GCM}
E_{ab} ={}&
  K_{ab} K^c{}_c
- K_a{}^c K_{bc}
+ \Rt_{ab} \\
&
- \frac{1}{2} \gamma_a{}^c \gamma_b{}^d \Rf_{cd}
- \frac{1}{2} \gamma_{ab} \gamma^{cd} \Rf_{cd}
+ \frac{1}{3} \gamma_{ab} \Rf \,. \nonumber
\end{align}
Here $\Rt_{ab}$ is the spatial 3-Ricci tensor while $\Rf_{ab}$ is the
4-Ricci tensor, and $K_{ab}$ is the extrinsic curvature of the
spacelike hypersurface. The second line of~\eqref{eq:E-from-GCM}
contains 4-Ricci terms which would vanish if the 4-metric was
Ricci-flat, for example if it solves the Einstein equations in vacuum.
These terms were not included in e.g.~\cite{Kidder2005}.

Meanwhile, for the magnetic Weyl tensor we
find the simple expression
\begin{align}
\label{eq:B-from-GCM}
B_{ab} = -\epsilon_{cd(a} D^c K_{b)}{}^d \,,
\end{align}
where $D_{a}$ is the covariant derivative induced on the 3-surface
which is compatible with the 3-metric, $D_{a}\gamma_{bc}=0$.

\onecolumngrid\newpage\twocolumngrid

\bibliography{References/References,References/dCS_paper}
\end{document}